\documentclass[PAPER, USenglish, cernpreprint, atlasdraft=false, coverpage=false, texmf, orcidlogo]{atlasdoc}
\usepackage{atlaspackage}
\usepackage{atlasbiblatex}

\usepackage{atlasphysics}
\graphicspath{{logos/}{figures/}}

\usepackage{ANA-TRIG-2022-02-PAPER-defs}

\usepackage{multirow}
\usepackage{booktabs}
\usepackage{xtab}
\usepackage{enumitem}
\usepackage{pdfpages}
\usepackage{rotating}
\usepackage{pdflscape}
\usepackage{tabulary}
\usepackage{lscape}
\usepackage{footmisc}
\usepackage{adjustbox}
\usepackage{float}

\AtlasTitle{Configuration, Performance, and Commissioning of the ATLAS $b$-jet
Triggers for the 2022 and 2023 LHC data-taking periods}

\AtlasAbstract{%
In 2022 and 2023, the Large Hadron Collider produced approximately two billion hadronic
interactions each second from bunches of protons that collide at a
rate of 40 MHz.
The ATLAS trigger system is used to reduce this rate to a few kHz for
recording.
Selections based on hadronic jets, their energy, and event topology reduce the
rate to $\mathcal O(10)$~kHz while maintaining high efficiencies for important
signatures resulting in $b$-quarks, but to reach the desired recording rate of
hundreds of Hz, additional real-time selections based on the identification of
jets containing $b$-hadrons ($b$-jets) are employed to achieve low thresholds
on the jet transverse momentum at the High-Level Trigger.
The configuration, commissioning, and performance of the real-time ATLAS
$b$-jet identification algorithms for the early LHC Run~3 collision data are
presented.
These recent developments provide substantial gains in signal efficiency for
critical signatures; for the Standard Model production of Higgs boson pairs, a
50\% improvement in selection efficiency is observed in final states with four
$b$-quarks or two $b$-quarks and two hadronically decaying $\tau$-leptons.
}

\AtlasRefCode{TRIG-2022-02}

\PreprintIdNumber{CERN-EP-2025-004}

\arXivId{2501.11420}

\AtlasJournalRef{JINST 20 (2025) P03002}
\AtlasDOI{10.1088/1748-0221/20/03/P03002}

\hypersetup{pdftitle={ATLAS document},pdfauthor={The ATLAS Collaboration}}

\begin{document}

\maketitle

\tableofcontents

%
%

\section{Introduction}

In its third proton--proton (\pp) run (Run 3), the Large Hadron Collider
(LHC)~\cite{Evans:2008zzb} produces collisions at $\sqrt{s} = 13.6$~\TeV every
25~ns. The average number of $pp$ collisions per bunch crossing, \pileupmu, delivered
to the ATLAS experiment~\cite{PERF-2007-01, GENR-2019-02} in the 2022 and 2023
data-taking has ranged from about 30 to about 70.
To contend with this 40~MHz event rate, the ATLAS experiment utilises a
two-staged triggering system~\cite{TRIG-2022-01}.
The first stage, composed of hardware-based selection algorithms
that run at the full 40~MHz rate, reduces this to a 100~kHz stream of events.
Software-based algorithms impose further selections to bring this rate down to
about 3~kHz for the event streams used in most analyses of the ATLAS data.
The hardware stage is known as the Level 1 (L1) trigger system, and the software
stage is called the High-Level Trigger (HLT) system.

Collimated final-state hadrons from the fragmentation of quarks and gluons
(``jets'') are produced at a high rate at the LHC, but jets containing \bhadrons
(\bjets) provide striking signatures in collider detectors that can be used to identify them~\cite{FTAG-2019-07}.
The identification of \bjets (\btagging) is a key component of a broad range of
ATLAS analyses of the LHC data and is used for example to probe properties of the
Higgs boson~\cite{HIGG-2018-04, HIGG-2018-13,HDBS-2019-29} and the top
quark~\cite{TOPQ-2018-17,TOPQ-2017-03} and to search for a wide variety
of possible processes beyond the Standard Model
(SM)~\cite{EXOT-2015-04,EXOT-2018-09,SUSY-2020-16}.

At a hadron collider experiment, the cross-section of multijet events is
substantial compared with electroweak or Higgs boson production
processes~\cite{Bellm:2019yyh}; analyzing hadronic final states therefore poses
a significant experimental challenge already at the trigger stage.
For fully hadronic final-states including $b$-quarks, \btagging is a strong tool
for reducing the trigger rate to the point of being manageable with the
available hardware and computing resources while maintaining a high signal
efficiency.
Several analyses of the LHC data probe processes with fully hadronic
final states that involve at least one \bjet.
These include Higgs boson and top-quark pair production events in the fully
hadronic channels~\cite{HDBS-2019-29,TOPQ-2018-18} and the production of new
resonances decaying preferentially to \bquarks~\cite{EXOT-2018-09}.

This article describes the real-time \bjet identification algorithms developed
by the ATLAS Collaboration for the LHC Run 3 $pp$ collision data,
highlighting differences relative to Run 2 ATLAS triggering strategy for
\bjets~\cite{TRIG-2018-08}.
First, an overview of the ATLAS detector and trigger systems is given.
The real-time \btagging algorithms utilised in the HLT are then introduced;
their optimisation and expected performance in simulation is also presented.
Comparisons between detector simulation and observed data collected during the
LHC 2022 and 2023 data-taking are shown, and triggering rates from this period
are reported.



\section{ATLAS detector}

The ATLAS detector at the LHC covers nearly the entire solid %
angle around the collision point.
It consists of an inner tracking detector surrounded by a thin superconducting
solenoid, electromagnetic and hadron calorimeters, and a muon spectrometer
incorporating three large superconducting air-core toroidal magnets.

The inner-detector system (ID) is immersed in a 2~T axial magnetic field and
provides charged-particle tracking in the range of $|\eta| < 2.5$.
\footnote{ATLAS uses a right-handed coordinate system with its origin at the
nominal interaction point (IP) in the centre of the detector and the $z$-axis
along the beam pipe.
The $x$-axis points from the IP to the centre of the LHC ring, and the $y$-axis
points upwards. Polar coordinates ($r$,$\phi$) are used in the transverse plane,
$\phi$ being the azimuthal angle around the $z$-axis.
The pseudorapidity is defined in terms of the polar angle $\theta$ as $\eta =
-\ln \tan(\theta/2)$. Angular distance is measured in units of $\Delta R \equiv
\sqrt{(\Delta\eta)^{2} + (\Delta\phi)^{2}}$.}
The high-granularity silicon pixel detector covers the interaction region and
typically provides four measurements per track, the first hit normally being in
the insertable B-layer (IBL).
It is followed by the silicon microstrip tracker (SCT), which usually provides
eight measurements per track.
These silicon detectors are complemented by the transition radiation tracker
(TRT), which enables radially extended track reconstruction up to $|\eta| = 2.0$.
The TRT also provides electron identification information based on the fraction
of hits (typically 30 in total) above a higher energy-deposit threshold
corresponding to transition radiation.

The calorimeter system covers the pseudorapidity range $|\eta| < 4.9$.
Within the region $|\eta| < 3.2$, electromagnetic calorimetry is provided by
barrel and endcap high-granularity lead/liquid-argon (LAr) calorimeters, with an
additional thin LAr presampler covering $|\eta| < 1.8$ to correct for energy
loss in material upstream of the calorimeters.
Hadron calorimetry is provided by the steel/scintillator-tile calorimeter (Tile
calorimeter), segmented into three barrel structures within $|\eta| < 1.7$, and
two copper/LAr hadron endcap calorimeters.
The solid angle coverage is completed with forward copper/LAr and tungsten/LAr
calorimeter modules optimised for electromagnetic and hadronic energy
measurements respectively.

The muon spectrometer (MS) comprises separate trigger and high-precision
tracking chambers measuring the deflection of muons in a magnetic field
generated by the superconducting air-core toroidal magnets.
The field integral of the toroids ranges between 2.0 and 6.0~Tm across most
of the detector.
Three layers of precision chambers, each consisting of layers of monitored drift
tubes, cover the region $|\eta| < 2.7$, except in the innermost layer of the
endcap region, where layers of small-strip thin-gap chambers and Micromegas
chambers both provide precision tracking in the region $1.3 < |\eta| < 2.7$.
The muon trigger system covers the range $|\eta| < 2.4$ with resistive-plate
chambers in the barrel, and thin-gap chambers in the endcap regions, and with
these small-strip thin-gap chambers and Micromegas chambers in the
innermost layer of the endcap.

The Run~3 detector configuration benefits from several upgrades compared with
that for Run 2 to maintain high detector performance at the higher pile-up levels
of Run 3.
The improvements include a new innermost layer of the muon spectrometer
in the endcap region, which provides higher redundancy and a strong reduction
in fake-muon triggers.
Other updates and further details are provided in Ref.~\cite{GENR-2019-02}%

An extensive software suite~\cite{SOFT-2022-02} is used in data
simulation, in the reconstruction and analysis of real and simulated data, in
detector operations, and in the trigger and data acquisition systems of the
experiment.


\section{Datasets and simulated events}

The results presented in this paper use data from \pp collisions with a
centre-of-mass energy $\sqrts = 13.6$\,\TeV, collected during the first two
years of the Run~3 of the LHC, in 2022 and 2023.

Monte Carlo (MC) simulations of top-quark pairs (\ttbar) produced in \pp
collisions are used throughout this paper to provide a sample of simulated jets
resulting from $b$-, $c$-, and light-flavour quarks.
These simulated events are used for the optimisation of online \btagging
algorithms and to compare the simulated performance of said algorithms to
observed performance in collision data.
The production of \ttbar events was modelled using the
\POWHEGBOX[v2]~\cite{Frixione:2007nw,Nason:2004rx,Frixione:2007vw,Alioli:2010xd}
generator at next-to-leading-order (NLO) with the \NNPDF[3.0nlo]~\cite{Ball:2014uwa} parton distribution function (PDF) set
and the \hdamp parameter\footnote{The
\hdamp parameter is a resummation damping factor and one of the
parameters that controls the matching of \POWHEG matrix elements to
the parton shower and thus effectively regulates the
high-\pT radiation against which the \ttbar system recoils.} set
to 1.5\,\mtop~\cite{ATL-PHYS-PUB-2016-020}.
The events were interfaced to \PYTHIA[8.230]~\cite{Sjostrand:2014zea} to model
the parton shower, hadronisation, and underlying event, with parameters set
according to the A14 set of tuned parameters (“tune”)~\cite{ATL-PHYS-PUB-2014-021} and using the
\NNPDF[2.3lo] set of PDFs~\cite{Ball:2012cx}.
This combination of calculations and tune parameters was found to provide
the best agreement to collision data in measurements of \btagging efficiency~\cite{FTAG-2018-01} and $b$-quark
fragmentation in top-quark decays~\cite{TOPQ-2017-19}.

Samples of simulated Higgs boson pair production (``di-Higgs boson production'') via
gluon--gluon fusion in the fully hadronic $bbbb$ and $bb\tau\tau$ final states are
used to assess the expected performance of the trigger chains that make use of
\btagging algorithms at the HLT\@.
These samples are generated with \POWHEGBOX[v2], the \PDFforLHC[15nlo] PDF
set~\cite{Butterworth:2015oua}, and \Pythia[8.244] to model the parton shower
hadronisation and underlying event, with parameters set according to the A14
tune.

The decays of bottom and charm hadrons were performed by
\EVTGEN[1.6.0]~\cite{Lange:2001uf}.
All simulated events have additional overlaid minimum-bias interactions
generated with \Pythia[8.160]~\cite{Sjostrand:2007gs} with the A3 set of tuned
parameters~\cite{ATL-PHYS-PUB-2016-017} and \NNPDF[2.3lo] parton
distribution functions to simulate pile-up background.\footnote{Pile-up
interactions correspond to additional $pp$ collisions accompanying the
hard-scatter $pp$ interaction in proton bunch collisions at the LHC.}
The simulated events were processed through the full ATLAS detector
simulation~\cite{SOFT-2010-01} based on
\GEANT~\cite{Agostinelli:2002hh}.


\section{The ATLAS trigger system}
\label{sec:trigger}

ATLAS records data from LHC collisions using a two-stage triggering system; the
first-level, L1, is custom hardware-based while the second-level, HLT, is
software-based.
The Data Acquisition (DAQ) system~\cite{TRIG-2022-01} transports data from
custom subdetector electronics through to offline processing, according to the
decisions made by the trigger.

The L1 trigger uses custom electronics to trigger on reduced-granularity
information from the calorimeter and muon detectors.
The L1 calorimeter (L1Calo) trigger takes signals from the calorimeter detectors
as input.
The Run-2 L1Calo system was operated in parallel to the upgraded L1Calo system
during the start of Run 3.
While the upgraded system was in operation during this time, it was not used for
jet selections at Level 1; rather the ``legacy'' Run-2 system was used.
In the legacy system, analogue detector signals are digitised and calibrated by
the preprocessor and sent in parallel to the cluster processors (CP) and the jet/energy-sum processors (JEP).
The CP system identifies electron, photon, and $\tau$-lepton candidates above a
programmable threshold, and the JEP system identifies jet candidates and
produces global sums of total and missing transverse energy.

The L1 muon (L1Muon) trigger uses hits from resistive-plate chambers (RPCs) in
the barrel and thin-gap chambers (TGCs) in the endcap to coarsely determine
the muon candidate momentum~\cite{TRIG-2018-01}.
To reduce the rate in the endcap regions of particles not originating from the
interaction point, the L1Muon trigger in Run~2 applied coincidence requirements
between the outer TGC station and either the inner TGC stations or the Tile
calorimeter.
In Run~3, the replacement of the original small wheels by the new small wheels allow for a further rate reduction from good rate tolerance and improved
resolution.

The L1 topological processor (L1Topo) system takes input Trigger OBjects (TOBs)
containing kinematic information from the L1Calo and L1Muon systems and applies
topological selections.
New modules were designed for Run 3 to accommodate input from the new L1Calo and
L1Muon systems.
During 2022 and 2023, the Run~2 (legacy) L1Topo system ran in parallel with the
upgraded L1Topo system, although the legacy system did not process inputs from
the L1Muon system.

The L1 trigger decision is formed by the central trigger processor (CTP); in the
legacy system this is done combining information mainly from the L1Calo and muon
trigger processors, while the upgraded system receives the information from the
L1Muon trigger system through the muon-to-central Trigger Processor
Interface~\cite{8532707}, which was also upgraded for Run~3.
The muon information is then processed together with L1Topo and L1Calo systems'
outputs.
The total L1 trigger rate of accepting collision events has an upper limit of
about 100 kHz, determined by the rate at which the detector can be read out.

If accepted by the L1 trigger, events are then sent to the HLT\@.
Here, algorithms reconstruct the event at progressively higher levels of detail
than at L1, either in restricted regions-of-interest (\rois), which are
regions of the detector in which candidate trigger objects are identified
by the L1 trigger, or in the full event.
The ATLAS jet, \bjet, and missing-transverse-energy trigger algorithms run in
the full-event context, while electron, muon, $\tau$-lepton, and photon
identification usually run in \rois defined by the L1 system.

A typical real-time (``online'') reconstruction sequence makes use of dedicated
fast trigger algorithms to provide early rejection, followed by more precise and
more CPU-intensive algorithms that are similar or identical to those used for
offline reconstruction to make the final selection.
The HLT software is incorporated in the same software framework used
offline to reconstruct recorded events but runs on a dedicated
computing farm composed of about 90k multi-processor units.
Events accepted by the HLT selection are distributed to a server cluster, where
they are compressed and written to disk or tape for storage.
The physics output rate of the HLT during an ATLAS data-taking run with the
nominal Run-3 physics menu and LHC conditions is on average 3~kHz, excluding
streams used for detector calibrations, trigger-level
analyses~\cite{EXOT-2016-20} and other specialised applications.
Most physics \bjet trigger chains are part of the \textit{Main Stream}, and
promptly undergo offline reconstruction right after data-taking, while some
$HH$-dedicated chains are instead part of the \textit{Delayed Stream}, and are
reconstructed only when resources are available~\cite{TRIG-2022-01}.
The pipeline can maintain about an 8~GB/s rate of data to offline storage.


\section{Algorithms}
\label{sec:algorithms}

After a brief description of the pertinent L1 trigger selection and
the main inputs to HLT \bjet identification, this paper focuses on the
optimisation and performance of the \btagging algorithms used in the HLT.

\subsection{Level 1 trigger selections}

Since track finding requires reading out the ATLAS ID subsystems,
which is only possible at a frequency much lower than the 40~MHz LHC maximum
bunch-crossing rate, the L1 selection for \bjets is based exclusively on
information from the ATLAS calorimeter and muon systems~\cite{TRIG-2022-01}.
In 2022 and 2023, the legacy L1 system introduced in Section~\ref{sec:trigger}
was used for the hardware jet selection.
Hadronic jets are reconstructed in the L1Calo systems as clusters of energy in
the grid of trigger towers (0.2 $\times$ 0.2 in $\eta$ and $\phi$) presenting
hadronic and electromagnetic transverse energy sum substantially above the
expectation from noise~\cite{TRIG-2016-01}.
Suppression of backgrounds from different bunch crossings (``out-of-time backgrounds’’) and pile-up from the same bunch crossing is improved
through noise thresholds that are $\eta$ dependent and through an energy
pedestal subtraction that varies based on the LHC bunch crossing being analysed.
Jets reconstructed in the calorimeter systems are used to select events that are
likely to contain high-momentum jets and to reject the large
background of diffractive and soft-QCD processes.

Since $b$-hadron decay chains include muons about 20\% of the
time~\cite{Zyla:2020zbs}, selecting events with muon candidates in addition to
calorimeter jets greatly reduces the rate of non-\bjet backgrounds with a
non-negligible \bjet selection efficiency, especially in the case of
multi-\bjet final states where the combinatorial probability of having at least
one muon from $b$-hadron decays grows quickly with the number of \bjets.
Some ATLAS trigger selections take advantage of this, either at L1 alone
(through the L1Topo system) or in both the L1 and HLT selections; specifically,
muon triggers are used for signatures targeting four or more \bjets and
for single-jet triggers used in \btagging calibrations.

\subsection{Key inputs to HLT \bjet identification}

Real-time software \btagging depends on three main reconstructed inputs:
hadronic jets, charged-particle tracks (tracks), and primary interaction
vertices (PVs).
A more detailed description of each of these ingredients is given in Section~4
of Ref.~\cite{FTAG-2019-07} for the ATLAS offline \btagging algorithms; here the
focus is on the primary differences between the inputs to offline and online
\btagging.

\subsubsection{Charged-particle tracking}

Charged-particle tracks are reconstructed in the ID~\cite{ATL-PHYS-PUB-2015-051}, and
several track reconstruction strategies are used, depending on the rate at which
they run.
Here is presented an overview of these strategies as pertains to the selection
of \bjets in the HLT, but a more complete description can be found in
Ref.~\cite{ATL-DAQ-PUB-2023-002} and in Section~5.1 of Ref.~\cite{TRIG-2022-01}.

A \emph{fast track finder} (FTF) -- less CPU intensive than
alternative strategies -- performs track pattern recognition and a fast track
fit.
To reduce CPU consumption, TRT hits are not included in the track fit in the FTF
stage.
A \emph{precision track finder}, also referred to as ``precision tracking,'' is
optionally run after the FTF stage.
The precision tracking strategy uses the output of FTF tracking, namely the
spacepoints assigned to tracks, and later runs the full offline track
fit on these spacepoints~\cite{IDTR-2022-04}.
This achieves higher CPU efficiency than using the full offline track
reconstruction algorithm.
Using the offline track fit in the precision tracking strategy provides
well-measured tracks with high purity after quality requirements and good
resolution relative to the full offline
tracks~\cite{IDTR-2022-04,TRIG-2022-01}.

Tracking may be run over the entire detector (full-scan tracking), in regions of
interest of a defined geometry around some seeding object (\roi tracking), or by
combining several \rois into a single larger \roi (super-\roi tracking).
Performing tracking in a limited subset of the detector reduces the computing
time required.
The choice of full-scan tracking, \roi tracking, or super-\roi tracking depends
on the event rate at which tracking is required and the performance needs of
selections that depend on these tracks.
In HLT selections involving \btagging, an instance of full-scan tracking using
the FTF strategy is used for primary-vertex reconstruction and jet finding;
tracking parameters are chosen to enable it to run very quickly over the full
detector but at the expense of the tracking efficiency.
Charged-particle tracks that are input to the \btagging algorithms themselves
are the result of FTF and precision tracking steps that run in jet \rois.
For some signatures involving multiple \bjets, the FTF full-scan tracking is
unaffordable at the input rate required; in such cases, a \btagging
``preselection'' is imposed based on a separate FTF reconstruction instance
operating in a super-\roi built from calorimeter jets.
In events with at least four calorimeter jets with transverse momentum $\pt > 20$~\GeV, full-scan
tracking requires about 1~second per event, super-\roi tracking about 250~ms per
event, and \roi precision tracking for \btagging about 350~ms per event.

The precision and FTF tracking efficiencies for charged pions with $\pt >
4~\GeV$ ranges from 90\% for ${|\eta| < 1}$ to 70\% in the forward region (${2.3
< |\eta| < 2.5}$) of the detector; the efficiency for both tracking
algorithms falls very quickly for charged particles
with $\pt < 1~\GeV$~\cite{TRIG-2022-01}.
Additional selection criteria for reconstructed tracks are applied in the
\btagging algorithms to maintain a high efficiency for charged particles from
heavy-flavour hadron decays while rejecting tracks originating from pile-up
interactions.
For example, tracks with poor fit quality are discarded.
These additional selections are detailed in Sections~4 and~5 of
Ref.~\cite{FTAG-2019-07}.

\subsubsection{Primary vertex reconstruction}

The reconstruction of primary vertices for each event is crucial for
high-performance \btagging, since the measured location of the hard-scatter
collision (i.e.\ the collision in a bunch crossing resulting in the process of
interest) is used as a reference point for calculating track and vertex
displacements~\cite{ATL-PHYS-PUB-2019-015}.
For use in \btagging algorithms, a track's transverse and longitudinal impact
parameters relative to the reconstructed PV, \dzero and \zzero, are
respectively defined as the track's distance of closest approach to the PV in
the transverse plane and the longitudinal separation between the PV and the
point on the track where \dzero is measured; this \dzero measuring point is
called the \enquote{perigee}.
The impact parameters \dzero and \zzero tend to be larger for tracks from
$b$-hadron decays than for those originating directly from the hard-scatter
interaction.

During LHC Run 3, PV finding in the HLT is performed using a Gaussian
distribution track-density seed finder, which seeds vertex-finding locations
based on reconstructed tracks, followed by an adaptive multi-vertex finder
algorithm~\cite{ATL-PHYS-PUB-2019-015} that associates tracks to vertex
candidates via association weights optimised through an annealing process.
For $pp$ collisions with a high track multiplicity, the vertex $z$ position
resolution is about $30~\mu$m, while the transverse position resolution in the
core of the distribution is 10 to $12~\mu$m.
For comparison, the luminous region typically is of the order of tens of
millimeters in $z$ and less than $10~\mu$m in the transverse plane.

\subsubsection{Jet reconstruction algorithms}

Two different types of hadronic jets are reconstructed in the HLT for the
purpose of \bjet identification; these differ primarily in their
input constituents, and the choice of constituent depends on the event rate at
which they can be constructed given computing constraints.
The first type of jet is built from topological clusters of calorimeter cells
with significant energy, calibrated to the electromagnetic energy
scale~\cite{JETM-2018-05}; these are called \emtopo jets.
The second is known as \enquote{particle-flow} jets, since their constituents
are particle-flow objects (PFOs) inferred from information from both the ATLAS
calorimeters and ID\@.
In particular, particle-flow jet reconstruction can take advantage of the
superior resolution of particle tracking when low-\pt charged hadrons yield both
calorimeter deposits and a charged-particle track; the calibration of the PFO
candidates considers both calorimetry and tracking
information~\cite{PERF-2015-09}.
PFO candidates that are not matched to the hard-scatter vertex are omitted
from jet finding, reducing the impact of pile-up interactions on the jet
response.
However, particle-flow jet reconstruction comes at a much higher computational
cost due to its reliance on event-wide ID track finding to construct
constituents and to determine the hard-scatter vertex.

The anti-$k_t$ algorithm with radius parameter $R=0.4$~\cite{Cacciari:2008gp},
implemented in \textsc{FastJet}~\cite{Fastjet}, is used for both collections of
jets.
Jets with $\pt < 20~\GeV$ or $|\eta| \ge 2.5$ are not considered for \btagging
for several reasons: (1) jets outside this $\eta$ range are beyond the
fiducial volume of the ATLAS ID, (2) the number of jets that must be considered
at low-$\pt$ quickly becomes computationally prohibitive, and (3) the
efficiency calibration of low-$\pt$ jets is extremely
challenging~\cite{TRIG-2012-01}.

To reduce the number of jets with large energy fractions from pile-up
collision vertices before \btagging algorithms are run, the \enquote{jet vertex
tagger} (JVT) algorithm is used~\cite{PERF-2014-03} for particle-flow
jets.
The JVT procedure is based on a multivariate discriminant for each jet within
$|\eta| < 2.4$ built from the ID tracks ghost-associated\footnote{The
ghost-association algorithm collects tracks within a jet's geometric catchment
area, taking into account possibly overlapping jet areas and is detailed in Ref~\cite{Cacciari:2008gp}.} with the
jet; in particular, jets with a large fraction of high-momentum tracks from
pile-up vertices are less likely to satisfy the JVT requirement.
The JVT efficiency for jets originating from the hard $pp$ scattering is
above 90\% in the simulation across this \pt range and grows with jet \pt.
Since the rate of pile-up jets with $\pt \ge 60~\GeV$ is sufficiently small, the
JVT requirement is removed above this threshold.

\subsubsection{Track-to-jet matching and jet labeling}

Tracks are matched to jets by setting a maximum allowed angular separation
$\Delta R$ between the track momenta, defined at the perigee, and the jet axis,
which is defined as the direction of the four-momentum sum of the jet
constituents.
Given that the decay products from higher-\pt\ $b$-hadrons are more collimated,
the $\Delta R$ requirement varies as a function of jet \pt, being wider for
low-\pt\ jets (0.45 for jet $\pt=20~\GeV$) and narrower for high-\pt\ jets (0.26
for jet $\pt=150~\GeV$); if more than one jet fulfils the matching criteria, the
closest jet is preferred~\cite{FTAG-2019-07}.
The jet axis is also used to assign signed impact parameters to tracks, where
the sign is defined to be positive if the track intersects the jet axis in the
transverse plane in front of the primary vertex, and negative if the
intersection lies behind the primary vertex~\cite{PERF-2012-04}.

The flavour of a jet in simulation is determined by the nature of the hadrons it
contains.
Jets are labelled as $b$-jets if at least one weakly decaying $b$-hadron having
$\pt \ge 5~\GeV$ is found within a cone of size $\Delta R=0.3$ around the jet
axis.
If no $b$-hadrons are found, $c$-hadrons and then $\tau$-leptons are searched
for, based on the same selection criteria.
The jets matched to a $c$-hadron ($\tau$-lepton) are labelled as $c$-jets
($\tau$-lepton jets).
The remaining jets are labelled as light-flavour jets.
For jets with more than one heavy-flavour hadron, e.g.\ from
gluon splitting into $b\bar b$ or $c\bar c$, the procedure above is still
followed, and $b\bar b$ ($c \bar c$) jets will receive a $b$ ($c$) label.

\subsection{Low-level identification algorithms}

The outputs of three ``low-level'' \bjet identification algorithms are used as
inputs to the full ``high-level'' HLT discriminants used for the trigger
decision.
These can be broadly categorised as secondary-vertex (SV) finders and neural-network-based discriminators with tracks as their inputs.

Two secondary-vertex finders were used in the HLT in 2022 data-taking, the
\jetfitter and \ssvf algorithms; each algorithm attempts to reconstruct the
\bhadron decay vertex and, in the case of \jetfitter, the subsequent \chadron
decay as a tertiary vertex.
The \ssvf vertexing algorithm finds at most one displaced vertex consistent with
a heavy-flavour hadron decay based on charged-particle tracks within a
jet~\cite{ATL-PHYS-PUB-2017-011}.
\jetfitter is a topological multi-vertex finding algorithm that attempts to
reconstruct the full \bhadron decay chain, employing a modified Kalman filter to
fit the \bhadron decay hypothesis~\cite{ATL-PHYS-PUB-2018-025}.
Many observables from \ssvf and \jetfitter vertex finding, including the
track multiplicity, invariant mass, and three-dimensional decay length
significance of the vertex candidates, are constructed for use as inputs to
high-level \btagging algorithms.

Features from a neural-network discriminator taking ID tracks as input are also
constructed for HLT \btagging.
This discriminator is based on the \texttt{DeepSet}
architecture~\cite{zaheer2017deep}, which is a universal approximator of
permutation-invariant functions with variable-length input sets.
The Deep Impact Parameter Sets (\dips) discriminator uses ten quantities for
individual tracks to determine the probability that a given jet
is a \bjet, \cjet, or light-flavour jet.
This follows the algorithm originally designed for offline
flavour-tagging~\cite{ATL-PHYS-PUB-2020-014,TRIG-2022-03}, but a subset of the
input variables used in the offline case are included: the transverse and
longitudinal impact parameters of the track relative to the primary vertex,
the corresponding signed impact parameter significances~\footnote{The impact
parameter significance of a track is defined as the ratio of the track impact
parameter to its uncertainty.}, the number of hits in the IBL, the number of hits
in the first pixel layer beyond the IBL, the total number of hits in the pixel
system, the total number of hits in the SCT, the $\Delta R$ between the track
and the jet axis, and the ratio of the track \pt to the jet \pt.
Using the notation of Ref.~\cite{ATL-PHYS-PUB-2020-014}, the per-track embedding
network $\phi$ comprises three layers containing 100, 100,
and 128 nodes each; the jet-wide network $F$ has four internal layers with 100,
100, 100, and 30 nodes each.
The resulting flavour probabilities are used as low-level inputs to high-level
\btagging algorithms.

\subsection{High-level taggers: \dloned and \gnone}

Two high-level \btagging algorithms are used for HLT \bjet identification:
\dloned and \gnone.
Both are constructed from neural networks and optimised using the categorical
cross-entropy loss to derive an optimal discriminant between $b$-, $c$-, and
light-flavour jets.

\dloned is a multi-layer perceptron (MLP) classifier that takes as inputs the
jet \pt and $|\eta|$ and several outputs from low-level taggers and emits
approximate probabilities that a jet has a given flavour, denoted $p_u$, $p_c$,
and $p_b$ for light-flavour, $c$-, and $b$-jet probabilities.
From \ssvf, key inputs are the invariant mass of the SV ($m_{SV}$), the
energy fraction of SV tracks ($E_{frac}$), the SV track multiplicity
($n_{trk}^{SV}$), the 3D and transverse distances ($L_{xyz}$ and $L_{xy}$)
between the PV and SV, the significance of the 3D distance between PV and SV
($S_{xyz}$), and the $\Delta R$ between the jet axis and the displacement of the
SV relative to the PV\@.
Similarly, for \jetfitter the most important inputs are the $b$-hadron SV
$m_{SV}$, $E_{frac}$, $L_{xy}$, $L_{xyz}$, $S_{xyz}$, $\Delta R$ to the jet
axis, and additionally the invariant mass and energy fraction of the possible
tertiary vertex.
Finally, the \dips classification scores, $p_u^{\dips}$, $p_c^{\dips}$, and
$p_b^{\dips}$, are used as inputs to \dloned.
The network architecture comprises eight dense layers of [256, 128, 60,
48, 36, 24, 12, 6] nodes per layer and the \texttt{ReLU} activation
function~\cite{ramachandran2017searchingactivationfunctions}.
For a complete list of inputs and network architectural details, see
Ref.~\cite{FTAG-2019-07}.
The \dloned algorithm was used for real-time \bjet identification in the HLT
during the 2022 data-taking.

During the 2023 LHC $pp$ collision run, a new flavour-tagging algorithm based on
graph neural networks, \gnone, was used~\cite{ATL-PHYS-PUB-2022-027}.
In addition to the jet transverse momentum and pseudorapidity, \gnone directly
uses about 20 quantities for each track associated with a jet rather than the
outputs of low-level taggers to discriminate between jet flavours.
Only tracks with $\pt > 500$~\MeV and satisfying loose requirements on the number
of associated silicon hits in the pixel and SCT systems are considered.
Among the 20 quantities, important input features include the track $q/p$, its
angular distance to the jet axis, its impact parameters relative to the
primary vertex, the corresponding impact parameter significances, and summary
quality criteria; Ref.~\cite{ATL-PHYS-PUB-2022-027} provides a detailed
description of all tracking quantities used.

In this article, the  \dloner tagging algorithm, based on the same architecture
as \dloned, but utilizing a low-level algorithm based on Recurrent Neural
Networks~\cite{ATL-PHYS-PUB-2017-003} instead of \dips, and used in a wide array
of physics results as an offline tagging algorithm~\cite{FTAG-2019-07}, is shown for performance comparisons only.
While \dloner was not deployed as part of the ATLAS trigger system in LHC Run 3,
its performance is shown in what follows to provide a perspective on the
algorithm evolution leading to the \dloned and \gnone algorithms.

\subsection{Classifier training procedure}

Following the prescription detailed in Ref.~\cite{FTAG-2019-07}, a training
sample is built of jets taken from simulated \ttbar production with at least one
leptonically decaying $W$ boson in the final state.
This yields a good mixture of $b$-, $c$-, and light-flavour jets with which to
train the neural-network classifiers.
This population of jets is resampled such that all flavours have the same
distribution in the two-dimensional $\pt \times |\eta|$ space to ensure that
these jet kinematic quantities are not used for discrimination; the $b$-jet
distribution is used as the resampling target.
The classifiers are trained using the \textsc{Adam}
optimizer~\cite{kingma2017adam} within the
\textsc{Keras}~\cite{chollet2015keras} and
\textsc{pytorch}~\cite{paszke2017automatic} frameworks.
The categorical cross-entropy loss is minimized to derive the optimal classifier
of $b$-, $c$-, and light-flavour jets.

In the online environment, the performance of flavour-tagging for relatively
low-\pt jets is particularly important:
the cross-section of multijet production falls quickly with the energy scale of
the event, to the point that above $\sim$400~\GeV, flavour-tagging is not needed
to reduce HLT rates to a tractable level.
As such, the online \btagging algorithms are trained only using jets
from \ttbar\ production without additional jets from high-$Q^2$
processes, which were included in previous studies for offline
flavour-tagging~\cite{FTAG-2019-07}.

\begin{figure}[t]
\centering
\subfloat[]{\adjustbox{width=.44\columnwidth}{\includegraphics{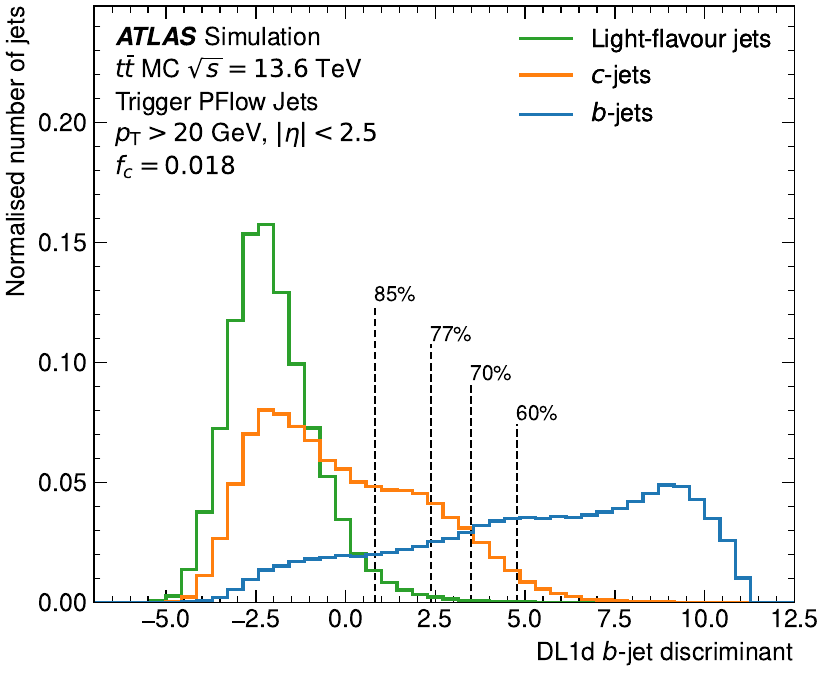}}}
\subfloat[]{\adjustbox{width=.44\columnwidth}{\includegraphics{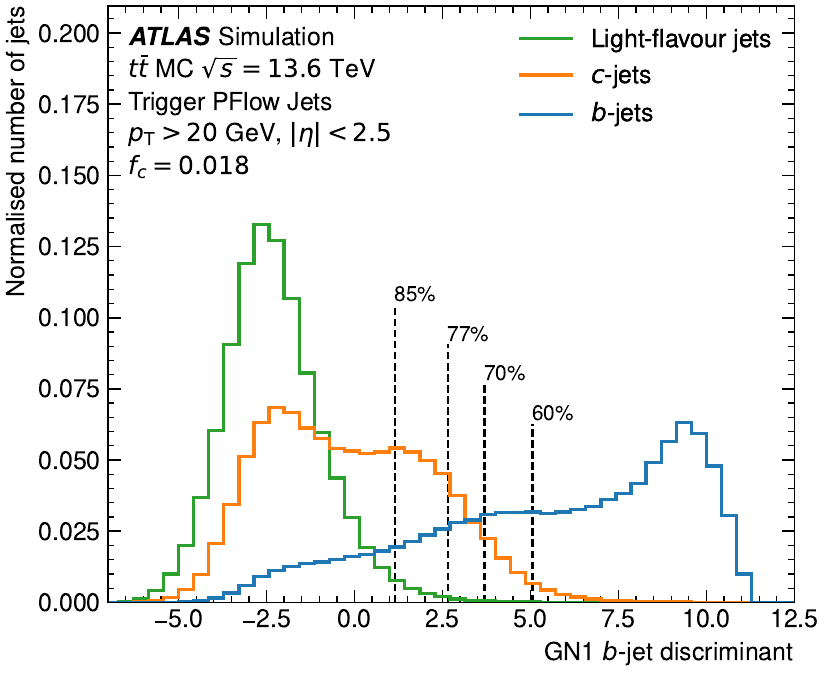}}}
\caption{The (a) \dloned and (b) \gnone discriminant scores for $b$-,
$c$-, and light-flavour jets in \ttbar events with $\pt > 20$~\GeV and $|\eta|
< 2.5$.
Dashed vertical lines indicate the discriminant selections for operating
points that are commonly used in the HLT.
}
\label{fig:tagscores}
\end{figure}

\subsection{Performance in simulation}

After training, a discriminant for \bjet identification is constructed from the
outputs of the \dloned and \gnone classifiers, which are the probabilities that
a jet belongs to the $b$-, $c$-, or light-jet categories.
For \bjet identification, this is defined as
\begin{equation}
D_b = \log \frac{p_b}{f_c p_b + (1 - f_c) p_u},
\end{equation}
where $f_c$ is a hyperparameter denoting the \cjet fraction of the
background population of jets.
For both \dloned and \gnone, the choice of $f_c = 0.018$ is made, following the
optimisation over many measurements and searches~\cite{FTAG-2019-07}, in
order to maximize the overlap between online and offline \btagging selections.
From these discriminants, operating points (OPs) are defined; e.g. the ``70\%
\btagging operating point'' is the value of $D_b$ for which 70\% of \bjets
with $\pt > 20$~\GeV in a SM \ttbar sample have a higher
discriminant score.
Figure~\ref{fig:tagscores} shows the real-time \btagging discriminant output for
$b$-, $c$-, and light-jets in a simulated sample of \ttbar events; clear
discrimination among the jet flavour classes is observed.

\begin{figure}[t]
\centering
\subfloat[]{\adjustbox{width=.44\columnwidth}{\includegraphics{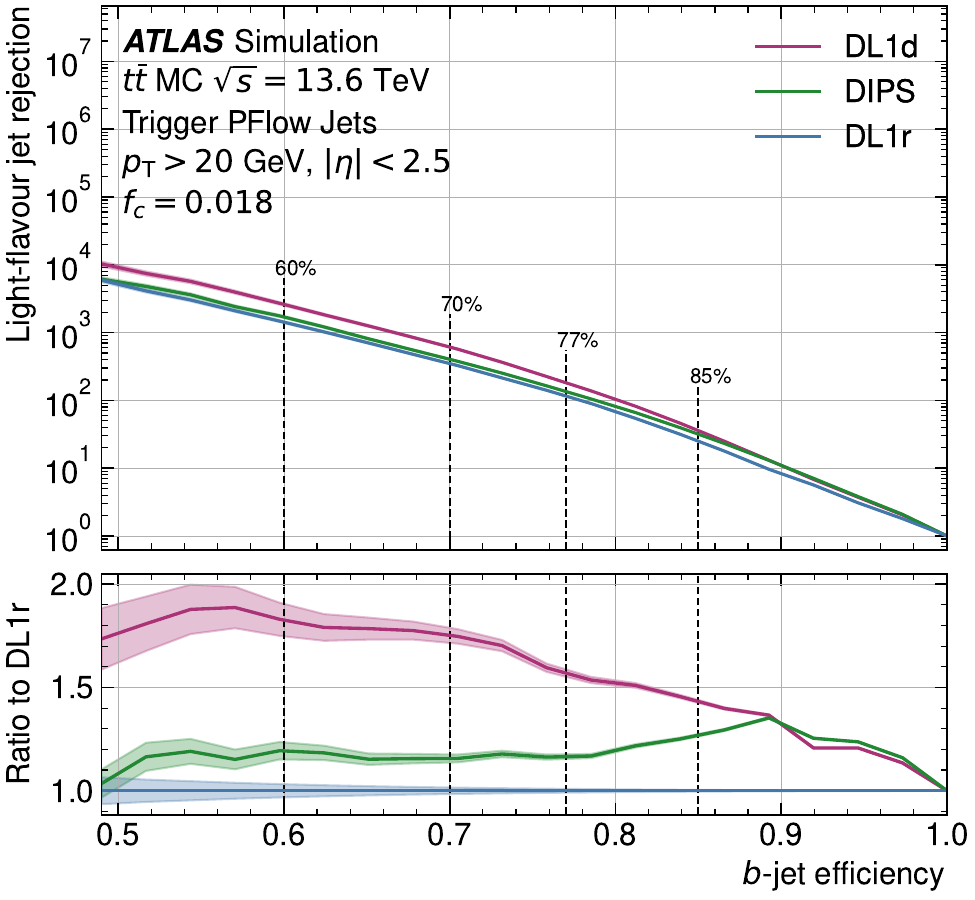}}}
\subfloat[]{\adjustbox{width=.44\columnwidth}{\includegraphics{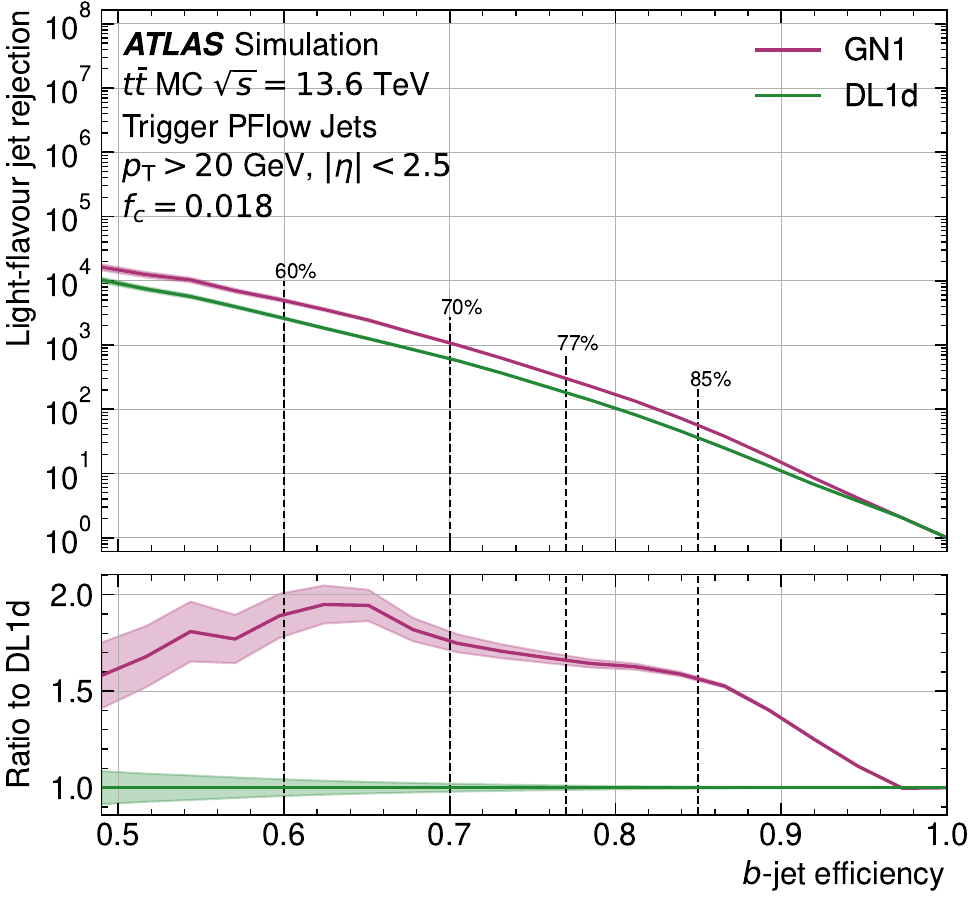}}}
\caption{The ROC curves for various flavour-tagging discriminants
using light-flavour jets and $b$-jets from \ttbar events with $\pt > 20$~\GeV and
$|\eta| < 2.5$.
The \dloned and \dips discriminants are compared with the (a) \dloner
tagger used for \btagging with the ATLAS detector in the LHC Run 2, and (b)
the \gnone classifier is compared with the \dloned classifier.
The lower panels show the ratio of the light-jet
rejection factors to \dloner (a) and \dloned (b). The shaded area represents the statistical uncertainty.
}
\label{fig:roc}
\end{figure}

The receiver operating characteristic (ROC) curves for the $b$-jet vs light-jet
classification task are shown in Figure~\ref{fig:roc} (a) for \dips and
\dloned compared with the Run 2 baseline \bjet identification algorithm, \dloner;
here, the rejection of background jets for a selection is defined as
$1/\varepsilon$, where $\varepsilon$ is the probability of a background jet to satisfy the selection.
The focus in the trigger system is on light-jet
discrimination, as opposed to \cjet discrimination, as most multijet events at the LHC do not produce a high-momentum charm hadron.
The simulated performance of \gnone is compared with \dloned in
Figure~\ref{fig:roc} (a): the \gnone classifier outperforms
\dloned for all \bjet efficiency OPs, providing up to a
factor two improvement in light-flavour jet rejection at a fixed \btagging
selection efficiency.

\section{The $b$-jet trigger menu}

To record events, the objects defined in Section~\ref{sec:algorithms} are used in combination according to a set of sequential rules named \textit{trigger chains}. A chain consists of an L1 trigger item and a series of HLT algorithms organised into distinct steps that reconstruct physics objects and apply kinematic selections to them~\cite{TRIG-2022-01}.
The list of trigger chains is referred to as the \textit{trigger menu}.
A large variety of $b$-jet trigger chains were deployed since the start of
Run 3 and were optimised using simulated events and validated using the first
data recorded at the beginning of the Run~3 data-taking.

\subsection{Estimated rates for $b$-jet chains}

\begin{figure}[t]
\centering
\includegraphics[width=0.7\columnwidth]{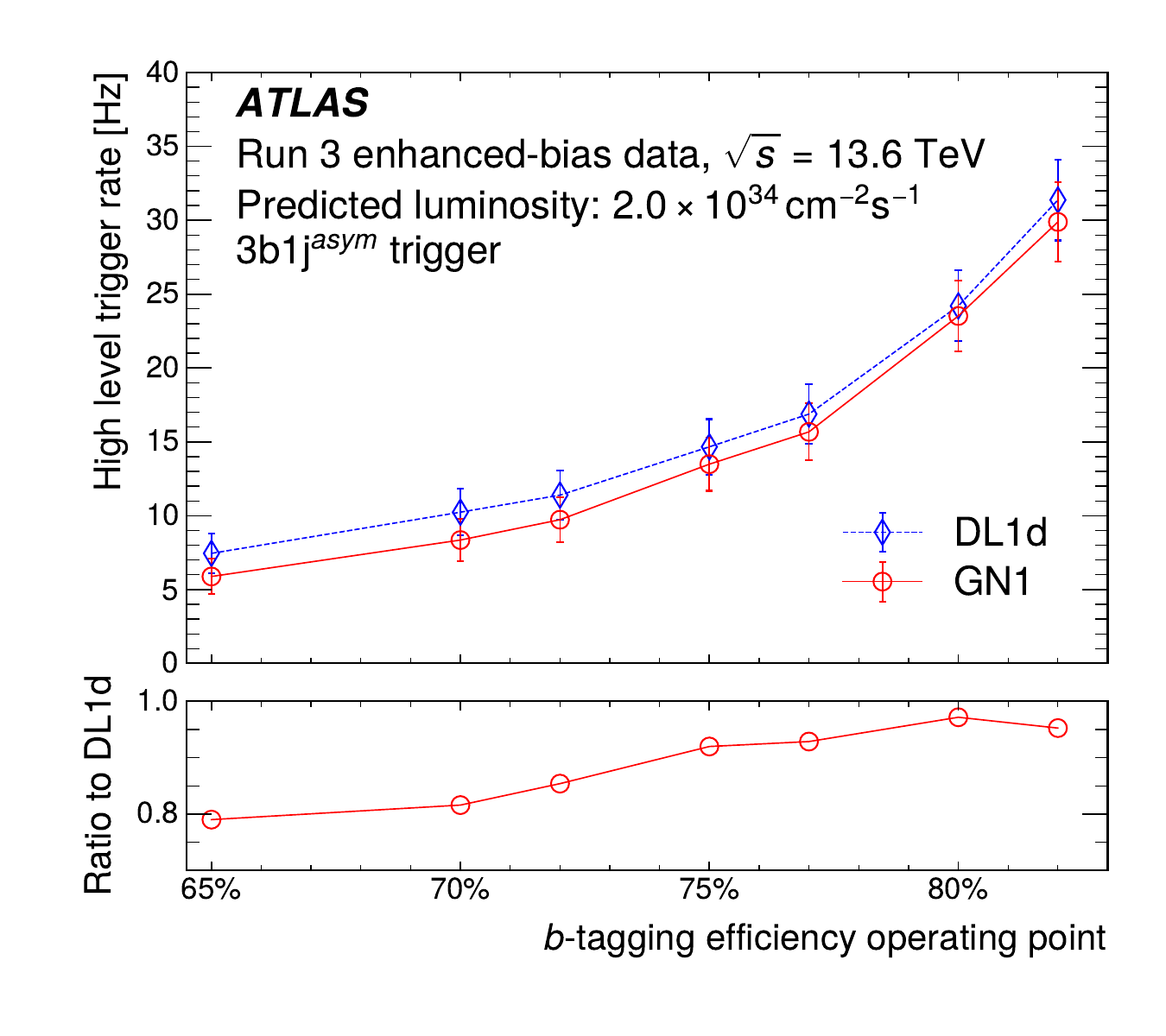}
\caption{
Estimated trigger rates by applying the trigger decision to enhanced
bias data for the multi-\bjet selection (\jbbbasym) using different \dloned
and \gnone HLT $b$-tagging working points.
}
\label{fig:3b1j-rates}
\end{figure}

Before data taking, rates for the proposed chains are evaluated with a
data-driven technique that uses a sample of collision events selected by the L1
system during previous runs to produce a compact data sample, referred to as
\textit{enhanced bias}~\cite{ATL-DAQ-PUB-2016-002}.
Enhanced bias data have the statistical power to assess the trigger rate of
algorithmic selections performed by the HLT\@.
This is a crucial asset when designing low-\pt $b$-jet trigger chains, where the
rates express a trade-off between requiring tighter $b$-tagging working points
or higher jet \pt thresholds at HLT to reject the overwhelming multijet
production.
Figure~\ref{fig:3b1j-rates} shows the estimated rate for a four-jet trigger with
at least three $b$-tagged jets (\jbbbasym), which is described in more
detail in Section~\ref{subsec:chains}) as a function of the working
point used for \dloned and for \gnone.
With an estimated \jbbbasym trigger rate of approximately 30 Hz, the
$\epsilon=82\%$ operating point was deployed online; \gnone outperforms
\dloned by about a 5\% relative rate at this particular operating point.

\subsection{List of $b$-jet chains}
\label{subsec:chains}

\begin{figure}[t]
\centering
\includegraphics[width=0.7\columnwidth]{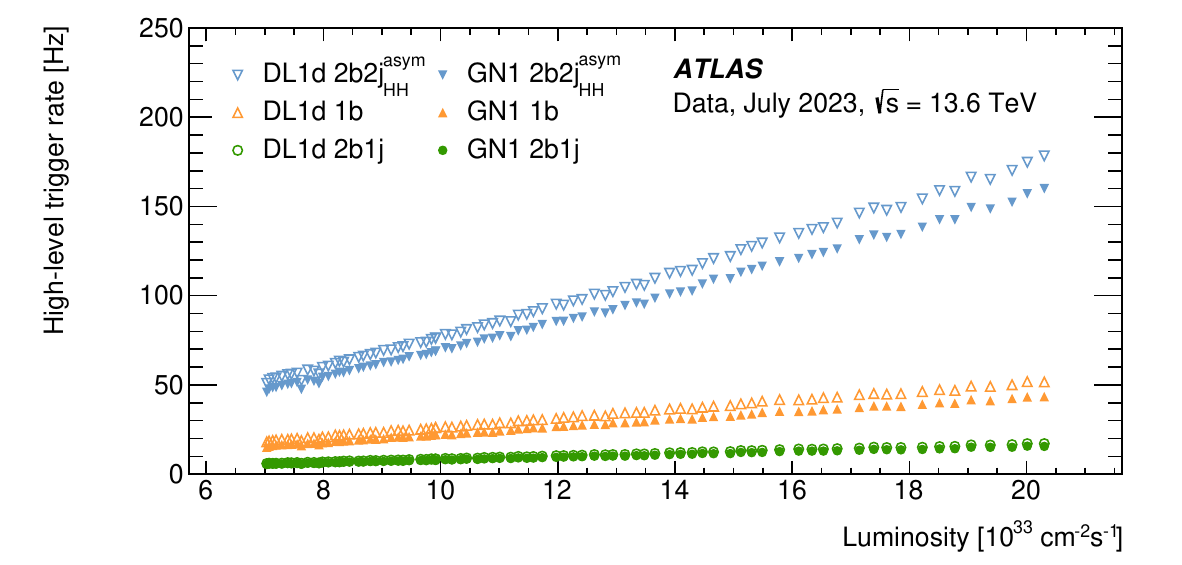}
\caption{ Output rates of selected ATLAS $b$-jet triggers as a function of the
instantaneous luminosity during 2023 proton--proton data taking with a
centre-of-mass energy of 13.6~\TeV and an LHC bunch-crossing interval of 25 ns for \dloned (empty markers) and \gnone (solid markers).
The trigger chains represented, \jjbbasym (inverted triangles), \monob
(triangles) and \jbb (circles)
are defined in
Tables~\ref{tab:menu_run3_unprescaled} and~\ref{tab:menu_run3_HH}.
}
\label{fig:rates}
\end{figure}

The list of $b$-jet trigger chains used in Run 3, itemized according to the requirements used at each level, from L1 to HLT preselection to HLT
is summarised in Table~\ref{tab:menu_run3_unprescaled} for the general-purpose chains, and in Tables~\ref{tab:menu_run3_HH},~\ref{tab:menu_run3_susy} and~\ref{tab:menu_run3_vbf} for the chains specialized for analyses such as di-Higgs boson~\cite{HDBS-2019-29}, supersymmetric searches~\cite{SUSY-2020-16} and Vector Boson Fusion (VBF) measurements.
Starting from Run 3, most $b$-jet trigger chains use a preselection step, to reduce the rate at which the precision tracking is performed. This consists of loose EMTopo jet $\pt$ requirements combined with a loose $b$-tagging preselection, based on the \dips algorithm which uses FTF as input~\cite{TRIG-2022-03}, referred to as \fd. The loose \fd working points deployed at preselection level present a $b$-tagging efficiency of 85\%, 90\% and 95\% when estimated on $t\bar{t}$ simulated events, and a light-jet rejection of 10, 5 and 2.5 respectively~\cite{TRIG-2022-03}.
Figure~\ref{fig:rates} shows the linear dependency of the output rates for the high-\pt threshold \monob, \jbb and for the \jjbbasym (seeded by three low-\pt jets at L1) $b$-jet triggers, as a function of the instantaneous luminosity for a collision run taken in 2022.

\begin{table}[H]
\centering
\renewcommand{\arraystretch}{1.2}
\caption{Details of the lowest-threshold \bjet triggers chains included in the Main Stream.
For L1 (HLT preselection and selection), $n \times \ET \; (\pt)$ indicates that $n$ trigger objects must satisfy the listed requirements, the L1 energy scale corresponds to an uncorrected EM scale~\cite{ATLAS:2020osu}.
In the $b$-tag column, $n \times b$ indicates that $n$ jets with $\pt > 20~\GeV$ must satisfy the specified \btagging operating point; these \bjet candidates need not be a subset of the jets satisfying the \pt-threshold requirements.
L1 and HLT jets are required to be within $|\eta| < 3.2$ and $|\eta| < 2.9$ respectively.
Unless otherwise specified, the preselection follows the $|\eta|$
requirements of the HLT\@.
Jets with $|\eta| > 2.5$ are not considered for $b$-tagging.
At HLT the value associated with the $b$-tagging efficiency ($\eff$) is the \dloned (\gnone) working point deployed in 2022 (2023) at HLT, while, for the HLT preselection level, it is the \fd working point.
For each trigger chain, the requirements listed in a given column must be satisfied independently.
The trigger thresholds denoted by the asterisk ($^{\ast}$) were moved down by $\sim$6\% from May 2023, thanks to an updated jet calibration at the HLT which improved the \pt response.
Jets in the HLT preselection are built from calorimeter inputs alone (\emtopo), while those in the HLT selection include charged-particle tracking (particle-flow).
\vspace{2mm}}
\begin{tabular}{ l l l l l  l }    \toprule
\multirow{2}{*}{Type}             & L1 threshold                           & \multicolumn{2}{l}{HLT preselection threshold}         &       \multicolumn{2}{l}{HLT selection threshold}  \\
& $\ET$ [\ensuremath{\text{G\electronvolt}}]                    & $\pt$ [\ensuremath{\text{G\electronvolt}}]       &   $b$-tag                &   $\pt$ [\ensuremath{\text{G\electronvolt}}]               &      $b$-tag \\    \toprule

\multirow{3}{*}{\monob}  & \multirow{3}{*}{1 $\times$ 100}              & 1 $\times$ 180    &            -             & 1 $\times$ $255^{\ast}$         &  1 $\times$ $b$, $\eff=70\%$   \\ \cmidrule(lr){3-6}
&                                              & 1 $\times$ 225    &            -             & 1 $\times$ $300^{\ast}$         &  1 $\times$ $b$, $\eff=77\%$    \\ \cmidrule(lr){3-6}
&                                              & 1 $\times$ 225    &            -             & 1 $\times$ $360^{\ast}$         &  1 $\times$ $b$, $\eff=85\%$  \\ \cmidrule(lr){1-6}

\multirow{2}{*}{\bb}             & \multirow{2}{*}{
1 $\times$ 100}                        & 1 $\times$ 140,   &  \multirow{2}{*}{2 $\times$ $b$, $\eff=85\%$}  & 1 $\times$ 175, & \multirow{2}{*}{2 $\times$ $b$, $\eff=60\%$} \\
&                                      & 1 $\times$ 45     &                                                & 1 $\times$ 60   &    \\ \cmidrule(lr){1-6}

\multirow{2}{*}{\jbb}            & 1 $\times$ 85,                       & 1 $\times$ 80,     & \multirow{2}{*}{2 $\times$ $b$, $\eff=90\%$}  & 1 $\times$ $150^{\ast}$, & \multirow{2}{*}{2 $\times$ $b$, $\eff=70\%$} \\
& 2 $\times$ 30                        & 2 $\times$ 45     &                                                & 2 $\times$ $55^{\ast}$   &                  \\ \cmidrule(lr){1-6}

\multirow{2}{*}{\bbb}            & 3 $\times$ 35,                       & \multirow{2}{*}{3 $\times$ 45} &  \multirow{2}{*}{2 $\times$ $b$, $\eff=95\%$} & \multirow{2}{*}{3 $\times$ $65^{\ast}$} &  \multirow{2}{*}{3 $\times$ $b$, $\eff=77\%$} \\
&   $|\eta| < 2.3$                     &                                &                                               &                                       &               \\ \cmidrule(lr){1-6}

\multirow{2}{*}{\jjjb}           & \multirow{2}{*}{4 $\times$ 20}       & \multirow{2}{*}{4 $\times$ 50} & \multirow{2}{*}{1 $\times$ $b$, $\eff=85\%$} &  1 $\times$ $75^{\ast}$ & \multirow{2}{*}{1 $\times$ $b$, $\eff=60\%$} \\
&                                      &                                 &                                             &  3 $\times$ $75^{\ast}$, $|\eta| < 3.2$  & \\ \cmidrule(lr){1-6}

\multirow{2}{*}{$\jjbb_{\texttt{cent}}$}         & 4 $\times$ 15,                        & \multirow{2}{*}{4 $\times$ 25} & \multirow{2}{*}{2 $\times$ $b$, $\eff=85\%$} & \multirow{2}{*}{4 $\times$ 35, $|\eta| < 2.5$} & \multirow{2}{*}{2 $\times$ $b$, $\eff=60\%$} \\
& $|\eta| < 2.5$                        &                                &                                              &                                                & \\ \cmidrule(lr){1-6}

\multirow{2}{*}{\jjbb}          & 4 $\times$ 15,                        & \multirow{2}{*}{4 $\times$ 25} & \multirow{2}{*}{2 $\times$ $b$, $\eff=85\%$} & 2 $\times$ 45 & \multirow{2}{*}{2 $\times$ $b$, $\eff=60\%$} \\
& $|\eta| < 2.5$                        &                                &                                              & 2 $\times$ 45, $|\eta| < 3.2$                  & \\ \cmidrule(lr){1-6}

\multirow{2}{*}{\bbHT}           & $\HT$ > 150,                         &                                &                                              & 2 $\times$ 45 & \multirow{2}{*}{2 $\times$ $b$, $\eff=70\%$} \\
& $M_{\textrm{jj}}$ > 400                &                                &                                              & $M_{\textrm{jj}} > 700$, $\HT > 300^{\ast}$   &    \\ \cmidrule(lr){1-6}

\multirow{2}{*}{\jbbb}          & 4 $\times$ 15,                       & \multirow{2}{*}{4 $\times$ 25} & \multirow{2}{*}{2 $\times$ $b$, $\eff=85\%$} & 3 $\times$ 35 & \multirow{2}{*}{3 $\times$ $b$, $\eff=70\%$} \\
& $|\eta| < 2.5$                        &                                &                                              & 1 $\times$ 35, $|\eta| < 3.2$                  & \\ \cmidrule(lr){1-6}

\multirow{3}{*}{\bbbb}           & \multirow{3}{*}{\shortstack[l]{
4 $\times$ 15, \\
$|\eta| < 2.5$ }}
& \multirow{3}{*}{4 $\times$  25} & \multirow{3}{*}{4 $\times$ $b$, $\eff=95\%$} & \multirow{3}{*}{4 $\times$ 35}  & 2 $\times$ $b$, $\eff=85\%$  \\
&                    &                                 &                                              &                                 & 2 $\times$ $b$, $\eff=70\%$  \\  \cmidrule(lr){6-6}
&                    &                                 &                                              &                                 &  4 $\times$ $b$, $\eff=77\%$   \\ \cmidrule(lr){1-6}

\multirow{2}{*}{\jjjbb}       & 5 $\times$ 15,                       & \multirow{2}{*}{5 $\times$ 25} & \multirow{2}{*}{2 $\times$ $b$, $\eff=85\%$} & 3 $\times$ 35 & \multirow{2}{*}{2 $\times$ $b$, $\eff=60\%$} \\
& $|\eta| < 2.5$                       &                                &                                              & 2 $\times$ 35, $|\eta| < 3.2$                  & \\

\bottomrule

\end{tabular}
\label{tab:menu_run3_unprescaled}
\end{table}

\begin{table}[htb]
\centering
\renewcommand{\arraystretch}{1.2}
\caption{Details of the asymmetric \bjet triggers dedicated for Higgs boson physics included in the Main and Delayed ($\ast$) Streams.
For L1 (HLT preselection and selection), $n \times \ET \; (\pt)$ indicates that $n$ trigger objects must satisfy the listed requirements, the L1 energy scale corresponds to an uncorrected EM scale~\cite{ATLAS:2020osu}.
In the $b$-tag column, $n \times b$ indicates that $n$ jets with $\pt > 20~\GeV$ must satisfy the specified \btagging operating point; these \bjet candidates need not be a subset of the jets satisfying the \pt-threshold requirements.
L1 (j) and HLT jets are required to be within $|\eta| < 3.2$ and $|\eta| < 2.4$ respectively. Unless otherwise specified, the preselection follows the $|\eta|$ requirements of the HLT\@.
Jets with $|\eta| > 2.5$ are not considered for $b$-tagging.
At HLT the value associated with the $b$-tagging efficiency ($\eff$) is the \dloned (\gnone) working point deployed in 2022 (2023) at HLT, while, for the HLT preselection level, it is the \fd working point.
For each trigger chain, the requirements listed in a given column must be satisfied independently.
Jets in the HLT preselection are built from calorimeter inputs alone (\emtopo), while those in the HLT selection include charged-particle tracking (particle-flow). L1 muons require two- and three-stations coincidence respectively for the barrel ($|\eta|<1.05$) and the endcap $1.05<|\eta|<2.4$.
\vspace{2mm}}
\begin{tabular}{ l l l l l l }    \toprule
\multirow{2}{*}{Type}             & L1 threshold                   & \multicolumn{2}{l}{HLT preselection}         &       \multicolumn{2}{l}{HLT selection}  \\
& $\ET$ [\ensuremath{\text{G\electronvolt}}]                    & $\pt$ [\ensuremath{\text{G\electronvolt}}]       &   $b$-tag                &       $\pt$ [\ensuremath{\text{G\electronvolt}}]               &      $b$-tag \\    \toprule

\multirow{4}{*}{\shortstack[l]{\jbbbasym}} %

& j: 1 $\times$ 45,
& \multirow{4}{*}{4 $\times$ 20}
& \multirow{4}{*}{2 $\times$ $b$, $\eff=85\%$}
& 1 $\times$ 80
& \multirow{4}{*}{3 $\times$ $b$, $\eff=82\%$}   \\

& $|\eta| < 2.1$
&
&
& 1 $\times$ 55
& \\

& 2 $\times$ 15,
&
&
& 1 $\times$ 28
& \\

& $|\eta| < 2.5$
&
&
& 1 $\times$ 20
& \\
\cmidrule(lr){1-6}

\multirow{4}{*}{\shortstack[l]{\jjbbasym}($\ast$)} %

& j: 1 $\times$ 45,
& \multirow{4}{*}{4 $\times$ 20}
& \multirow{4}{*}{2 $\times$ $b$, $\eff=85\%$}
& 1 $\times$ 80
& \multirow{4}{*}{2 $\times$ $b$, $\eff=77\%$}   \\

& $|\eta| < 2.1$
&
&
& 1 $\times$ 55
& \\

& 2 $\times$ 15,
&
&
& 1 $\times$ 28
& \\

& $|\eta| < 2.5$
&
&
& 1 $\times$ 20
& \\
\cmidrule(lr){1-6}

\multirow{4}{*}{\shortstack[l]{$\mu$+\jjbbasym}($\ast$)} %
& \multirow{4}{*}{\shortstack[l]{j: 1 $\times$  20  \\ \\ \\ 1 $\times$ 15  \\ \\ \\ $\mu$: 1 $\times$ 8 ($\pt$)}}
& \multirow{4}{*}{4 $\times$ 20}
& \multirow{4}{*}{2 $\times$ $b$, $\eff=85\%$}
& 1 $\times$ 80
& \multirow{4}{*}{2 $\times$ $b$, $\eff=77\%$}   \\

&
&
&
& 1 $\times$ 55
& \\

&
&
&
& 1 $\times$ 28
& \\

&
&
&
& 1 $\times$ 20
& \\
\cmidrule(lr){1-6}

\end{tabular}
\label{tab:menu_run3_HH}
\end{table}

\begin{table}[htb]
\centering
\renewcommand{\arraystretch}{1.2}
\caption{Details of the missing transverse energy plus \bjet triggers dedicated for supersymmetric searches included in the Main Stream.
For L1 (HLT preselection and selection), $n \times \ET \; (\pt)$ indicates that $n$ trigger objects must satisfy the listed requirements, the L1 energy scale corresponds to an uncorrected EM scale~\cite{ATLAS:2020osu}.
In the $b$-tag column, $n \times b$ indicates that $n$ jets with $\pt > 20~\GeV$ must satisfy the specified \btagging operating point; these \bjet candidates need not be a subset of the jets satisfying the \pt-threshold requirements.
L1 (j) and HLT jets are required to be within $|\eta| < 3.2$ and $|\eta| < 2.9$ respectively. Unless otherwise specified, the preselection follows the $|\eta|$ requirements of the HLT\@. Jets with $|\eta| > 2.5$ are not considered for $b$-tagging.
At HLT the value associated with the $b$-tagging efficiency ($\eff$) is the \dloned (\gnone) working point deployed in 2022 (2023) at HLT, while, for the HLT preselection level, it is the \fd working point.
Cell-based missing transverse energy ($\ET^{\text{miss}}$) \cite{TRIG-2022-01} is used for early reduction, whereas the final $\ET^{\text{miss}}$ selection at HLT is based on PFOs.
For each trigger chain, the requirements listed in a given column must be satisfied independently.
Jets in the HLT preselection are built from calorimeter inputs alone
(\emtopo), while those in the HLT selection include charged-particle
tracking (particle-flow).
\vspace{2mm}}
\begin{tabular}{ l l l l l }       \toprule
Type                             & L1 threshold                         & HLT preselection                        &                   \multicolumn{2}{l}{HLT selection} \\
& $\ET$ [\ensuremath{\text{G\electronvolt}}]                          & $E^{\text{miss}}_{\text{T,cell}}$ [\ensuremath{\text{G\electronvolt}}]      &       $\pt$ [\ensuremath{\text{G\electronvolt}}]               &      $b$-tag \\ \toprule

\multirow{2}{*}{\shortstack[l]{\xeba}} %
&   $\ET^{\text{miss}} > 40$
& \multirow{2}{*}{60}
& \multirow{2}{*}{j: 1 $\times$  80}
& \multirow{2}{*}{ 1 $\times$ $b$, $\eff=60\%$}\\

& j: 2 $\times$ 50
&
&
& \\
\cmidrule(lr){1-5}

\multirow{2}{*}{\shortstack[l]{\xeb}} %
& \multirow{2}{*}{$\ET^{\text{miss}} > 55$}
& \multirow{2}{*}{50}
&  $\ET^{\text{miss}} > 85$
& \multirow{2}{*}{1 $\times$ $b$, $\eff=60\%$}\\

&
&
& j: 1 $\times$  100
& \\
\cmidrule(lr){1-5}

\multirow{2}{*}{\shortstack[l]{\xebb}} %
&   $\ET^{\text{miss}} > 55$
& \multirow{2}{*}{50}
&  $\ET^{\text{miss}} > 85$
& \multirow{2}{*}{2 $\times$ $b$, $\eff=60\%$} \\

& j: 2 $\times$ 15
&
&   j: 2 $\times$ 45
& \\
\cmidrule(lr){1-5}

\multirow{2}{*}{\shortstack[l]{\xebbb}} %
&   $\ET^{\text{miss}} > 40$
& \multirow{2}{*}{50}
&  $\ET^{\text{miss}} > 70$
& \multirow{2}{*}{3 $\times$ $b$, $\eff=60\%$} \\

& j: 3 $\times$ 15,  $|\eta| < 2.5$
&
&   j: 3 $\times$ 35
& \\

\bottomrule

\end{tabular}
\label{tab:menu_run3_susy}
\end{table}

\begin{table}[htb]
\centering
\renewcommand{\arraystretch}{1.2}
\caption{Details of the forward jets plus \bjet trigger chains dedicated to VBF measurements included in the Main Stream.
For L1 (HLT preselection and selection), $n \times \ET \; (\pt)$ indicates that $n$ trigger objects must satisfy the listed requirements, the L1 energy scale corresponds to an uncorrected EM scale~\cite{ATLAS:2020osu}.
In the $b$-tag column, $n \times b$ indicates that $n$ jets with $\pt > 20~\GeV$ must satisfy the specified \btagging operating point; these \bjet candidates need not be a subset of the jets satisfying the \pt-threshold requirements.
L1 (j) and HLT jets are required to be within $|\eta| < 3.2$ and $|\eta| < 2.9$ respectively,
while forward jets (f) are required to have $3.1 < |\eta| < 4.9$ both at L1 and HLT.
Unless otherwise specified, the preselection follows the $|\eta|$ requirements of the HLT\@.
Jets with $|\eta| > 2.5$ are not considered for $b$-tagging.
At HLT the value associated with the $b$-tagging efficiency ($\eff$) is the \dloned (\gnone) working point deployed in 2022 (2023) at HLT, while, for the HLT preselection level, it is the \fd working point.
For each trigger chain, the requirements listed in a given column must be satisfied independently.
Jets in the HLT preselection are built from calorimeter inputs alone
(\emtopo), while those in the HLT selection include charged-particle
tracking (particle-flow).
\vspace{2mm}}
\begin{tabular}{ l l l l l }       \toprule
Type                             & L1 threshold                         & HLT preselection                        &                   \multicolumn{2}{l}{HLT selection} \\
& $\ET$ [\ensuremath{\text{G\electronvolt}}]                          & $\pt$ [\ensuremath{\text{G\electronvolt}}]                             &            $\pt$ [\ensuremath{\text{G\electronvolt}}]           &      $b$-tag \\ \toprule

\multirow{2}{*}{\shortstack[l]{\bff}} %

& 1 $\times$ 25, $|\eta| < 2.3$
& \multirow{1}{*}{\shortstack[l]{1 $\times$ 45}}
& \multirow{1}{*}{1 $\times$ 55}
& \multirow{2}{*}{1 $\times$ $b$, $\eff=70\%$} \\

& 2 $\times$ 15, 3.1 $<|\eta|< 4.9$
& \multirow{1}{*}{\shortstack[l]{1 $\times$ 40}}
& \multirow{1}{*}{2 $\times$ 45, 3.1 $<|\eta|< 4.9$}  \\

\cmidrule(lr){1-5}

\multirow{3}{*}{\shortstack[l]{\bbf}} %

& 1 $\times$ 40,  $|\eta| < 2.5$
& 1 $\times$ 60
& 1 $\times$ 80
& 1 $\times$ $b$, $\eff=70\%$ \\

& 1 $\times$ 25
&  1 $\times$ 45
&  1 $\times$ 60
&   1 $\times$ $b$, $\eff=85\%$ \\

& 1 $\times$ 20, 3.1 $<|\eta|< 4.9$
& 1 $\times$ 40
& 1 $\times$ 45, 3.1 $<|\eta|< 4.9$
& \\

\cmidrule(lr){1-5}

\multirow{3}{*}{\shortstack[l]{\bjf}} %

& 1 $\times$ 40, $|\eta| < 2.5$
&1 $\times$ 60
& 1 $\times$ 80, $|\eta|<$ 2.4
& \multirow{3}{*}{\shortstack[l]{1 $\times$ $b$, $\eff=60\%$}} \\

& 1 $\times$ 25
&  1 $\times$ 45
& 1 $\times$ 60
& \\

& 1 $\times$ 20,  3.1 $<|\eta|< 4.9$
&  1 $\times$ 40
&  1 $\times$ 45, 3.1 $<|\eta|< 4.9$
& \\

\bottomrule

\end{tabular}
\label{tab:menu_run3_vbf}
\end{table}

\subsection{Performance in \ttbar\ enriched data}

Top quark pairs are copiously produced at the LHC and can be used as a control sample to compare data and simulations.
Since the branching fraction of the top-quark decay into a $W$ boson and a $b$-quark is nearly 100\%, identifying these events can provide a large sample of relatively pure $b$-jets, an important asset when assessing $b$-tagging performance for both offline~\cite{FTAG-2018-01} and online~\cite{TRIG-2018-08} algorithms.
Moreover, MC simulated $t\bar{t}$ events are used, before data-taking, to
optimise the algorithm working points which are later deployed online. Once
collected using lepton triggers, top quark pairs decay into final state with at
least one lepton approximately 56\% of the
time~\cite{ParticleDataGroup:2024cfk}; data enriched in $t\bar{t}$ events
provide an unbiased sample in which the $b$-jet trigger performance for $b$-jets
in data can be compared with MC\@.
This comparison is performed once data is processed through the offline reconstruction algorithms.
These algorithms benefit from the better knowledge of detector conditions, obtained during the offline processing, and are therefore more precise than their online equivalent.
Moreover, the amount of CPU time per event is less of a constraint for offline
reconstruction, whereas the online peformance is largely affected by the
limitations on the CPU consumption imposed to the FTF seed finding, a major
source of inefficiency for online tracking~\cite{PERF-2015-08}.
This has a direct impact on the better performance of the offline $b$-tagging algorithms relative to their online equivalent.

\subsubsection{Offline reconstruction}

Electron candidates are reconstructed from an isolated electromagnetic calorimeter energy deposit with $ |\eta_\mathrm{cluster}| < 2.47$, which is matched to a track in the ID~\cite{EGAM-2018-01}.
The electron track must satisfy $|z_0 \sin \theta| < \SI{0.5}{mm}$,  where $\theta$ is the track polar angle, and $|d_0|/\sigma_{d_0} < 5$, where $\sigma_{d_0}$ is the uncertainty in $d_0$.
For the tight likelihood identification working point~\cite{EGAM-2018-01} used in this work, the isolation criteria, defined as an upper requirement on the sum of the transverse energy or momentum reconstructed in a cone of size around the electron, excluding the energy of the electron itself, depends on the electron's \pt. Furthermore, electrons are required to have $\pt>4.5$~\GeV.

Particle-flow jets are constructed from offline PFOs analogously to online particle-flow jets introduced in Section~\ref{sec:algorithms}.
These jets are then calibrated to the particle level by applying a jet energy scale derived from simulation with in situ corrections based on collected data~\cite{JETM-2018-05}.
A cleaning procedure is used to identify and remove jets arising from calorimeter noise or
non-collision backgrounds. To suppress pile-up jets within $|\eta|<2.4$, a discriminant called the `jet vertex tagger'
(nnJVT) is constructed using a neural network method~\cite{PERF-2014-03}.
Jets within $|\eta|< 2.5$ are $b$-tagged using the \dloned $b$-tagging algorithm~\cite{FTAG-2018-01,FTAG-2019-07} with a working point corresponding to a $b$-tagging efficiency of 77\%,  measured in a sample of simulated \ttbar events.  The corresponding rejection factors are approximately 200 and 6 for light- and $c$-jets, respectively.

\subsubsection{Data Monte-Carlo comparison in $t\bar{t}$ enriched events}

Data enriched in $t\bar{t}$ events are selected with the following requirements:
\begin{itemize}
\item{Satisfy the isolated single-electron-plus-jets trigger, corresponding to an \et requirement of 26 \GeV for the electron and \pt $\geq$ 20 \GeV for the two jets at the HLT\@.}
\item{Contain an offline electron with $\ET \ge 28\,\GeV$, $|\eta| < 2.47$, excluding the transition region between the barrel and endcap cryostats ($1.37 \le |\eta| < 1.52$), satisfying the tight identification and isolation requirements.  }
\item{Exactly one electron satisfying the trigger selection.}
\item{At least four offline jets with $|\eta|< 2.5$ and with $\pt > 120\,\GeV$, for the leading jet, $\pt > 70\,\GeV$, for the subleading jet and $\pt > 30\,\GeV$, for any extra jet in the event.}
\item{At least two offline jets must satisfy the offline $b$-tagging \dloned algorithm criteria for the 77\% efficiency working point.}
\end{itemize}
After this selection, the data sample contains at least 90\% $t\bar{t}$ events,
and it is used to compare data and MC simulation~\cite{TOPQ-2020-09}.

\begin{figure}[htbp]
\centering
\subfloat[]{\adjustbox{width=.45\columnwidth}{\includegraphics{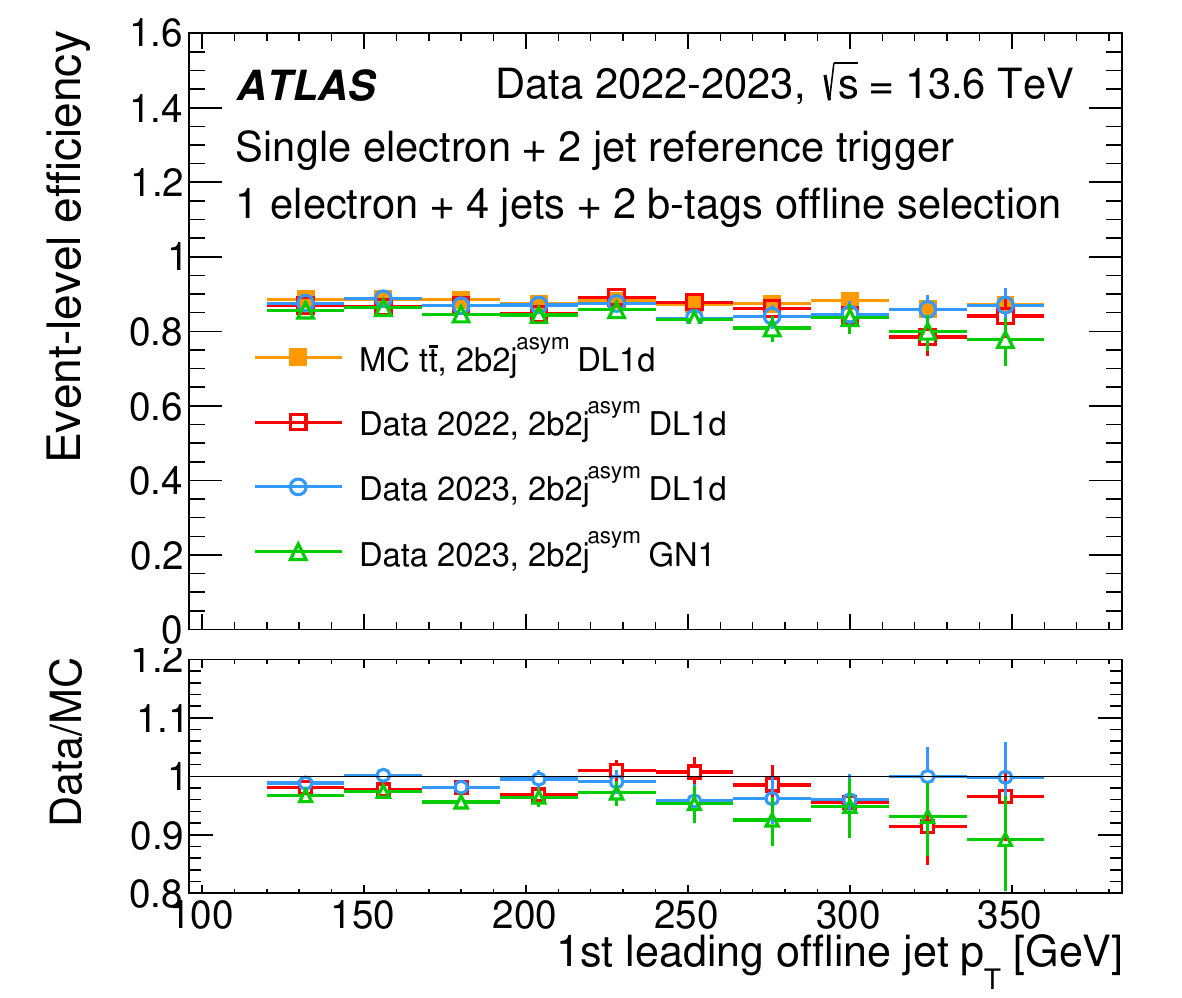}}}
\subfloat[]{\adjustbox{width=.45\columnwidth}{\includegraphics{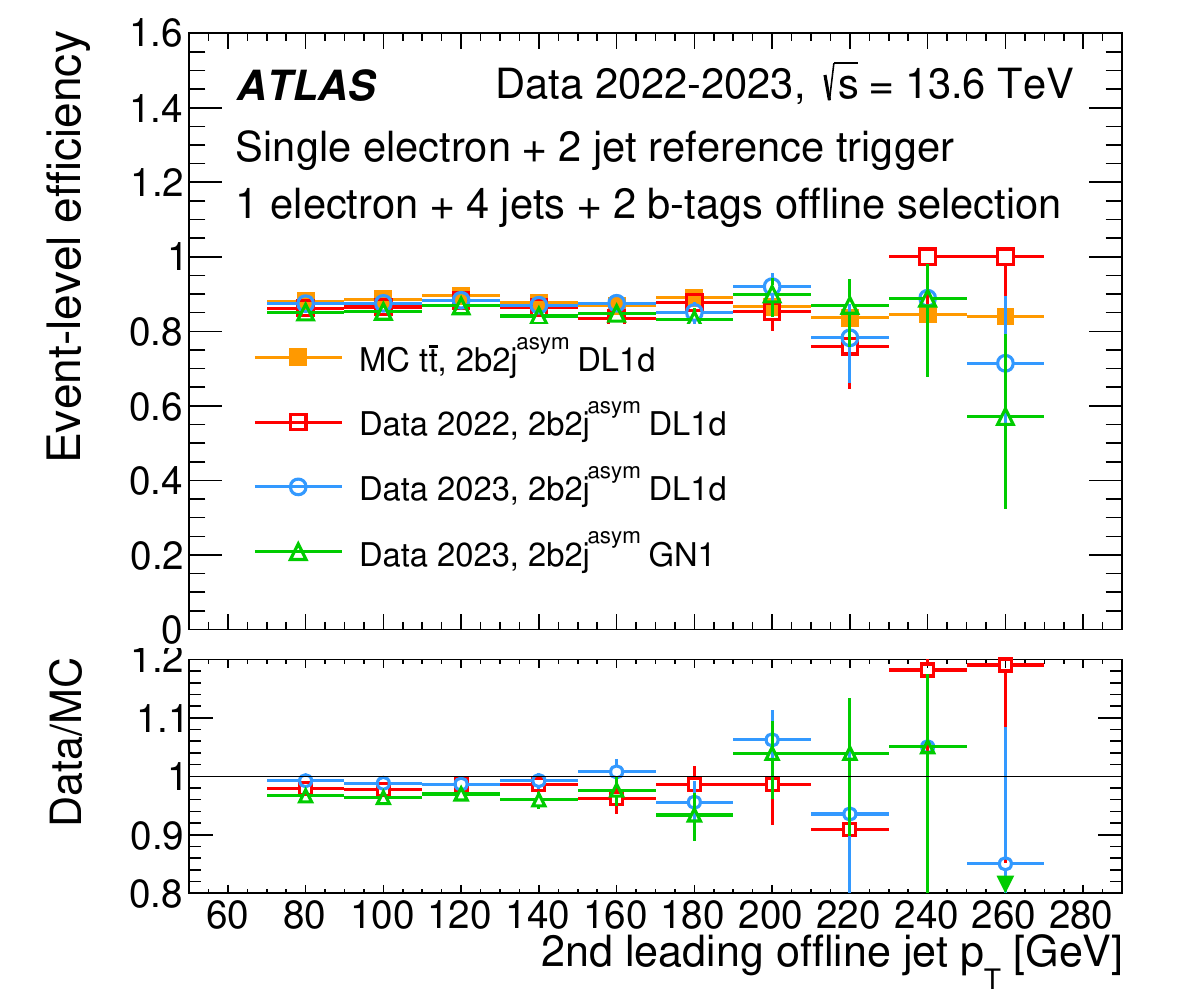}}} \\
\subfloat[]{\adjustbox{width=.45\columnwidth}{\includegraphics{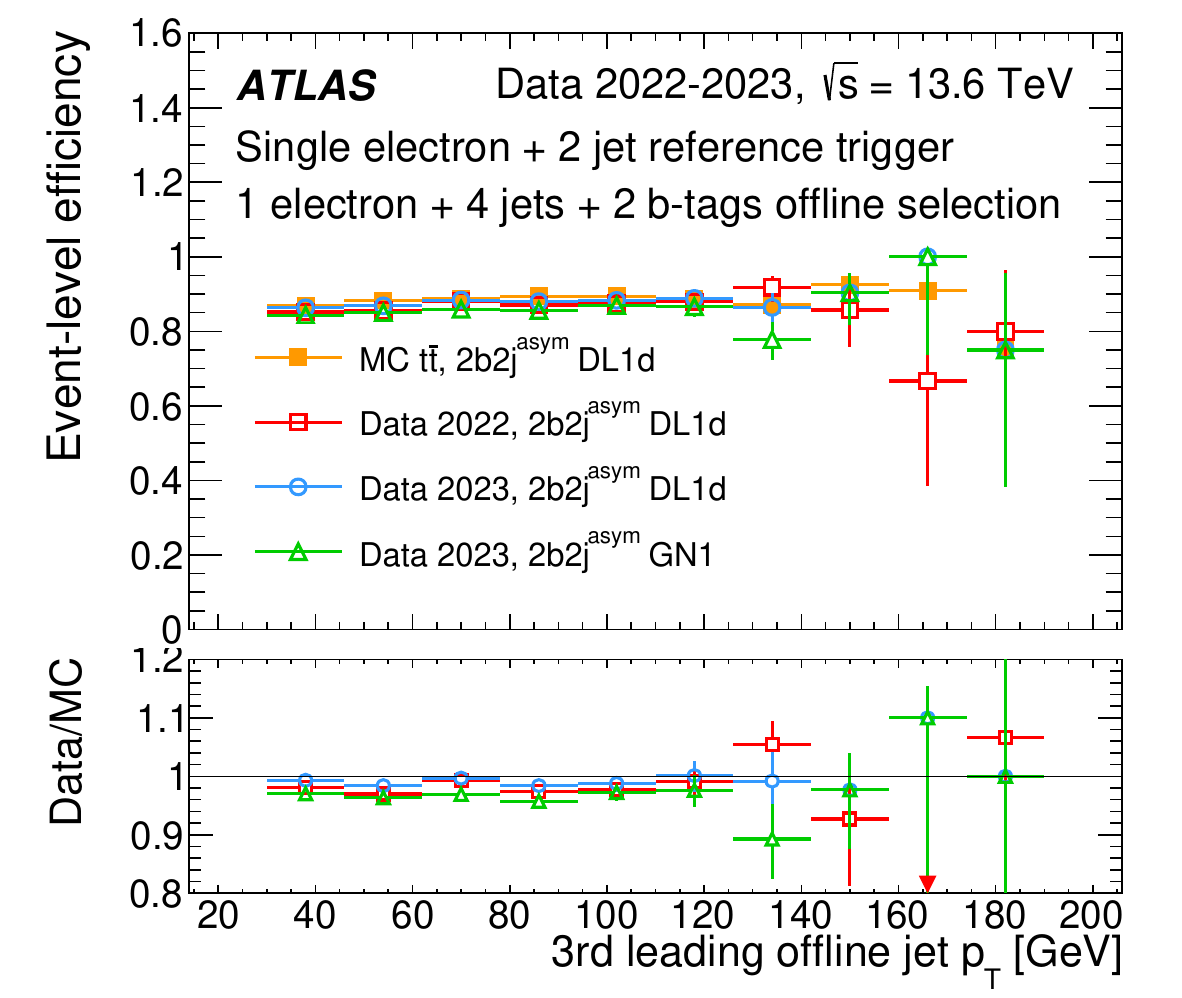}}}
\subfloat[]{\adjustbox{width=.45\columnwidth}{\includegraphics{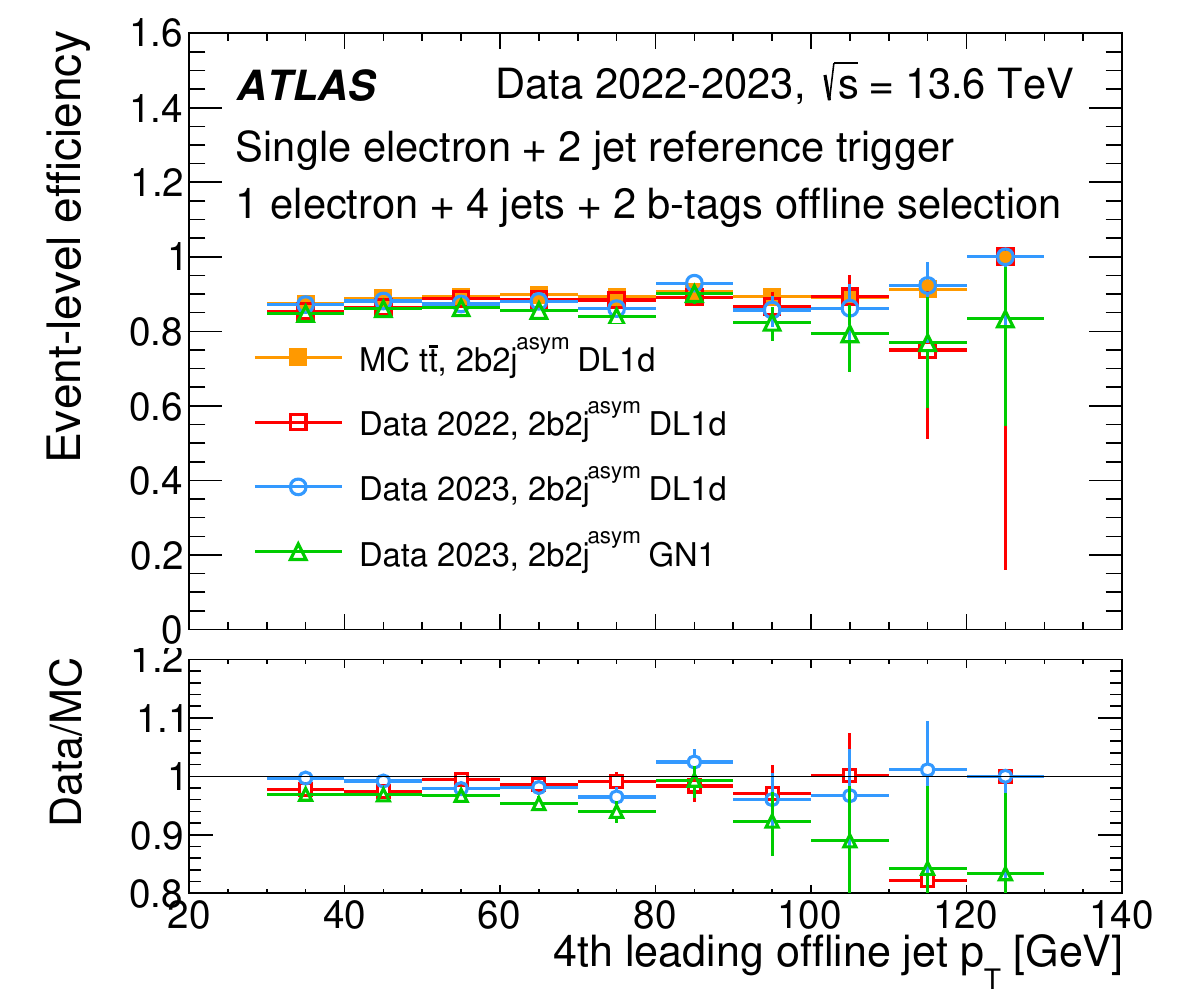}}}
\caption{
Trigger efficiencies in data and simulation for a \jjbbasym HLT selection after requiring an isolated-electron-plus-jet trigger and an offline selection requiring one electron, four jets and two $b$-tagged jets in the event.
The efficiency of the \jjbbasym trigger selection is shown relative to the \pt\ of the (a) leading, (b) subleading, (c) third leading  and (d) fourth leading reconstructed offline jet.
Efficiencies, shown above the plateau of the jet trigger turn-on, are sensitive to differences between $b$-jet performance in data and MC $t\bar{t}$ events. The lower panels show the ratio between data and the simulated $t\bar{t}$ events.
}
\label{fig:2b2j-effs}
\end{figure}

Figure~\ref{fig:2b2j-effs} shows the percentage of the selected lepton-triggered events satisfying the \jjbbasym $b$-jet trigger defined in Table~\ref{tab:menu_run3_HH}, for both \dloned and \gnone algorithms.
This \jjbbasym event-level trigger efficiency is shown independently as a function of the first-, second-, third- and fourth-leading offline jet \pt.
The offline jet \pt requirements are chosen such that at least 95\% of the events passing the  $t\bar{t}$ offline selection satisfy the L1 and HLT jet requirements of the \jjbbasym trigger
chain; this condition is referred to as ``above the plateau of the jet trigger turn-on''.
The jet \pt requirements depend on the presence of the isolated electron from the top-quark decay which can participate as a reconstructed jet in the \jjbbasym decision.
After the offline jet requirements, the \jjbbasym efficiency shown in Figure~\ref{fig:2b2j-effs} is only sensitive to residual difference between the performance of the $b$-jet trigger in data and MC $t\bar{t}$ events.
Finally, Figure~\ref{fig:2b2j-effs} shows an overall good agreement for the correlation of the offline and online $b$-tagging between data and MC simulation.

\subsection{Efficiency improvements for \hhbbbb and \hhbbtt compared with Run~2}

The trigger menu described in Section~\ref{subsec:chains} yields substantial
improvements in the data recording efficiency relative to Run~2 for two of dominant final
states of Higgs boson pair production: \hhbbbb and \hhbbtt.
The gain in efficiency is driven by improvements in the \bjet identification algorithms, described in Section~\ref{sec:algorithms}, that consequently allows loosening jet transverse momentum thresholds.

\begin{figure}[t]
\centering
\includegraphics[width=0.8\columnwidth]{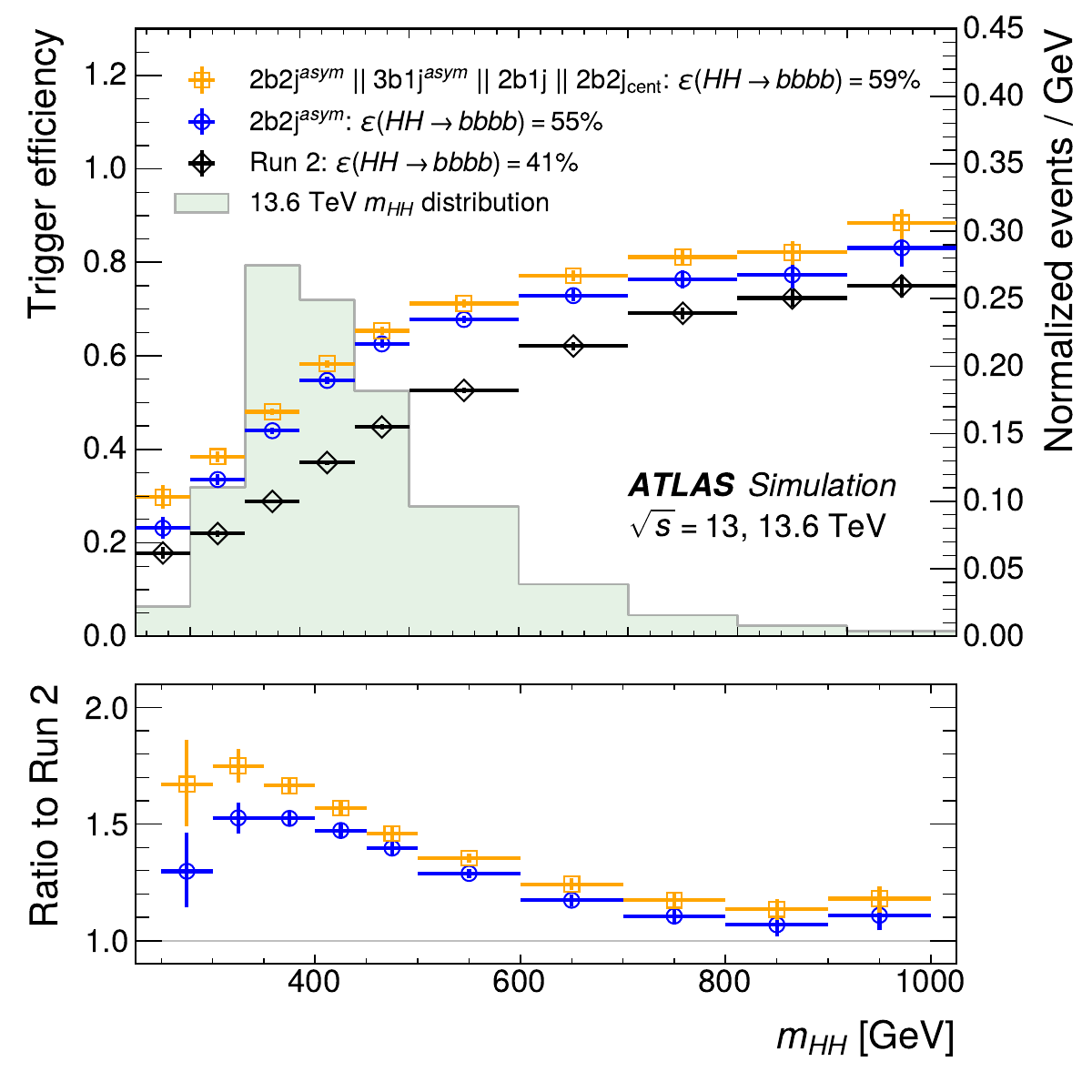}
\caption{
Performance of the \bjet trigger HLT selections in a simulated sample
of \hhbbbb events.
The trigger efficiency relative  to inclusive \hhbbbb events is shown as a
function of the true di-Higgs boson invariant mass, \mhh, for the \jjbbasym selection and
the union of \jjbbasym, \jbbbasym, \jbb, and $\jjbb_{\texttt{cent}}$ triggers, as introduced
in  Tables~\ref{tab:menu_run3_unprescaled} and~\ref{tab:menu_run3_HH},
compared with the trigger selection employed in the Run~2
analysis~\cite{HDBS-2019-29}.
The \jjbbasym trigger has the highest efficiency of any single trigger
chain in the Run~3 menu.
}
\label{fig:hh4beffs}
\end{figure}

An inclusive efficiency gain of about 50\% is achieved compared with the Run~2 trigger strategy for the SM \hhbbbb signal.
The efficiency improvements are especially large ($\sim 75\%$) in the low $HH$
invariant mass (\mhh) region close to the $2m_H$ threshold, which is known to
have particular sensitivity to the Higgs boson self-interaction
strength~\cite{GOUZEVITCH2020100039}.
Figure~\ref{fig:hh4beffs} compares the \hhbbbb signal efficiency of the Run~2 ATLAS trigger strategy with the new menu deployed in Run~3 as a function of \mhh.

\begin{figure}[t]
\centering
\includegraphics[width=0.8\columnwidth]{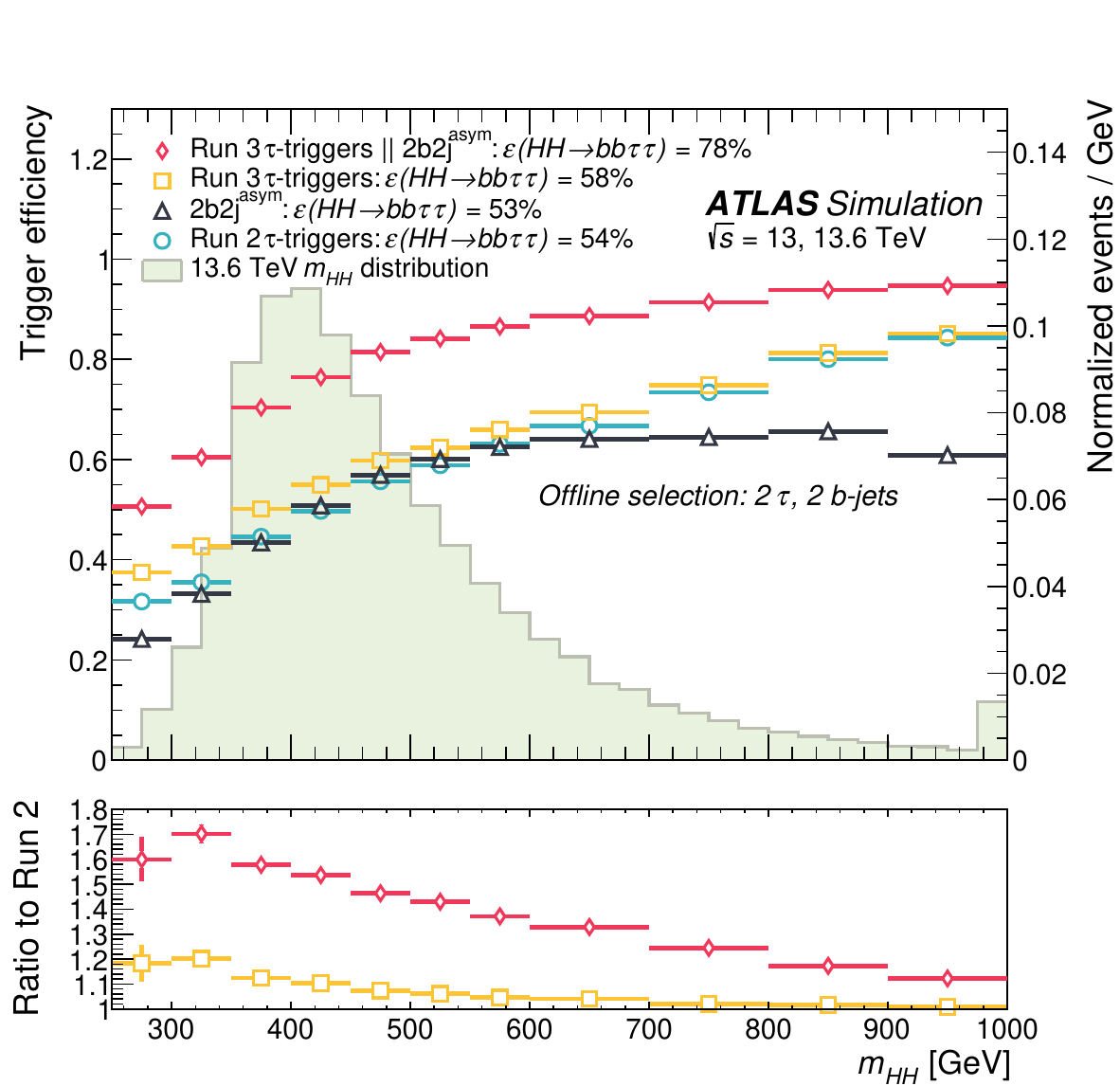}
\caption{
Performance of the \bjet trigger HLT selections in a simulated sample of
\hhbbtt events: the trigger efficiency as a function of the true di-Higgs
boson invariant mass, \mhh.
The \jjbbasym trigger selection compared with a trigger selection based purely
on HLT $\tau$-lepton identification is shown.
To ensure efficiencies are comparable across trigger strategies,
they are measured relative  to events within the detector acceptance for
the \hhbbtt analysis; these events must have two offline $\tau$-lepton candidates
matched to true hadronically decaying $\tau$-leptons and two offline
jets matched to \bhadrons.
The $\tau$-lepton based triggers shown follow the strategy used for the Run~2
\hhbbtt analysis~\cite{HDBS-2019-27} together with an additional di-$\tau$-lepton
trigger with lower \pt thresholds for Run~3, requiring at least one identified
$\tau$-lepton in the HLT system with $\pt > 30$~\GeV and a second with $\pt
> 20$~\GeV.
Efficiencies are calculated relative  to a selection of two reconstructed
\bjets ($\pt > 20$~\GeV) and two reconstructed $\tau$-leptons ($\ptvis > 25,
20$~\GeV) within the inner detector acceptance ($|\eta| < 2.5$).
}
\label{fig:hhbbtteffs}
\end{figure}

The \jjbbasym trigger chain also efficiently selects events from the
\hhbbtt process in which both $\tau$-leptons decay hadronically; for SM
\hhbbtt production, an efficiency above 50\% is predicted, in simulation, for events
passing the \hhbbtt fiducial selection: two
\bjets having $\pt > 20$~\GeV and two $\tau$-leptons having the hadronic component of the transverse momentum ($\ptvis$) $> 20$~\GeV, within the
inner detector acceptance ($|\eta| < 2.5$).
The \jjbbasym chain moreover achieves a large unique efficiency ($> 40$\%) with respect to a trigger strategy selecting events with pairs of $\tau$-leptons in
the HLT\@.
This strategy was used by the ATLAS Collaboration in Run~2 searches for the
\hhbbtt process~\cite{HDBS-2019-27}, where the leading-\pt (sub-leading-\pt)
$\tau$-lepton must have $\pt > 35$~\GeV ($\pt > 25$~\GeV).
In Run~3, ATLAS deployed a similar trigger chain with looser $\tau$-lepton \pt
thresholds at 30 and 20~\GeV.
A comparison between the trigger efficiencies for these various \hhbbtt trigger
strategies is shown differentially in \mhh in Figure~\ref{fig:hhbbtteffs}.
To ensure the efficiencies are comparable and relevant for the ATLAS search for
\hhbbtt production, they are calculated relative  to the \hhbbtt fiducial
selection.

\section{Offline and online $b$-jet trigger monitoring}

As the trigger is the first step in the ATLAS event selection of LHC data, its
behaviour must be understood at the deepest level since any failure during
data-taking is unrecoverable.
The quality of $b$-jet triggered data is continuously assessed, thanks to the
trigger system monitoring infrastructure~\cite{TRIG-2019-04}, which is part of
the ATLAS Data Quality Monitoring system~\cite{DAPR-2018-01}.
For the trigger, this consists of two parts: the online and the offline
monitoring systems.
The online monitoring system allows a real-time check of the quality of
collected events during the data-taking, and provides quick detection of
processing failures or reconstruction issues, ensuring that the ATLAS Trigger and Data Acquistion~\cite{TRIG-2022-01}
operates properly and collects high quality data.
Using the Data Quality Monitoring Display (DQMD), shifters in the ATLAS control
room can monitor distributions for $b$-tagging relevant quantities,
such as track impact parameters, properties of secondary vertices and jet
$b$-tagging weights, both inclusively, i.e. every time the algorithm is called,
and separately for specific trigger chains.
Inclusive quantities can provide high statistics distributions, available to the
shifter as histograms, although these are more sensitive to changes in the
trigger menu which could arise in the middle of the run from changes in the beam
conditions.
On the other hand, the exclusive monitoring of specific chains permits checking data behaviour in a stable environment, where each monitored trigger chain is chosen to have a different jet flavour composition, as an example multijet events are dominated by the light-jet component, while selecting data where muons are embedded in hadronic jets allows to enhance the $b$-jet component by taking advantage of the large semimuonic branching ratio of $B$-hadrons.

After the data are fully reconstructed, the offline monitoring allows a more
detailed quality assessment.
A set of automatic software routines run dedicated algorithms that first select events, based on the precise offline reconstructed physics objects,
compare relevant trigger and offline distributions with the ones taken in a reference run and then make it available to the offline shifter, via a web-based display.

\begin{figure}[htbp]
\centering
\subfloat[]{\adjustbox{width=.44\columnwidth, cfbox=OliveDrab 3pt 2pt}{\includegraphics{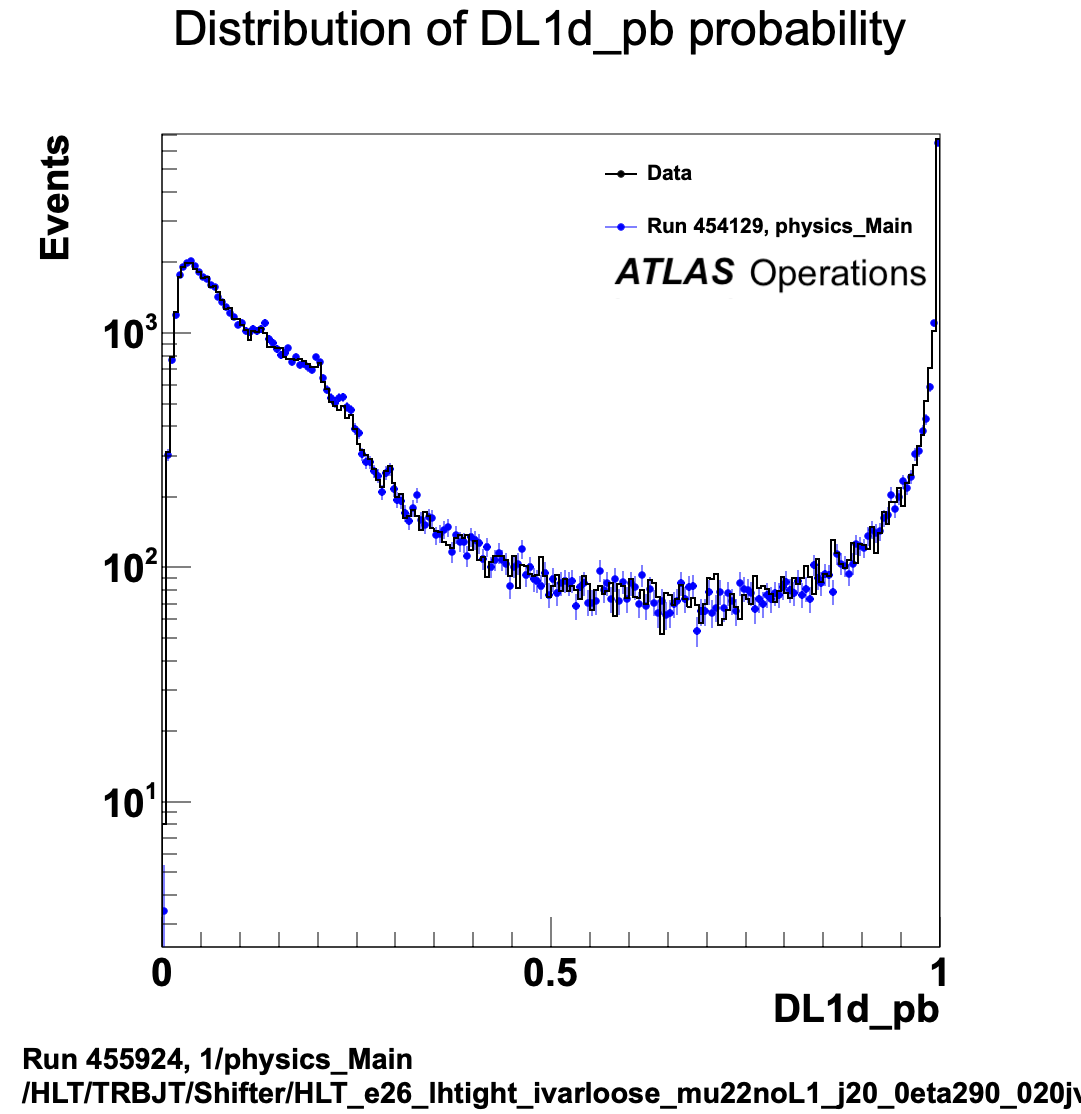}}}\hfill
\subfloat[]{\adjustbox{width=.44\columnwidth, cfbox=Gold 3pt 2pt}{\includegraphics{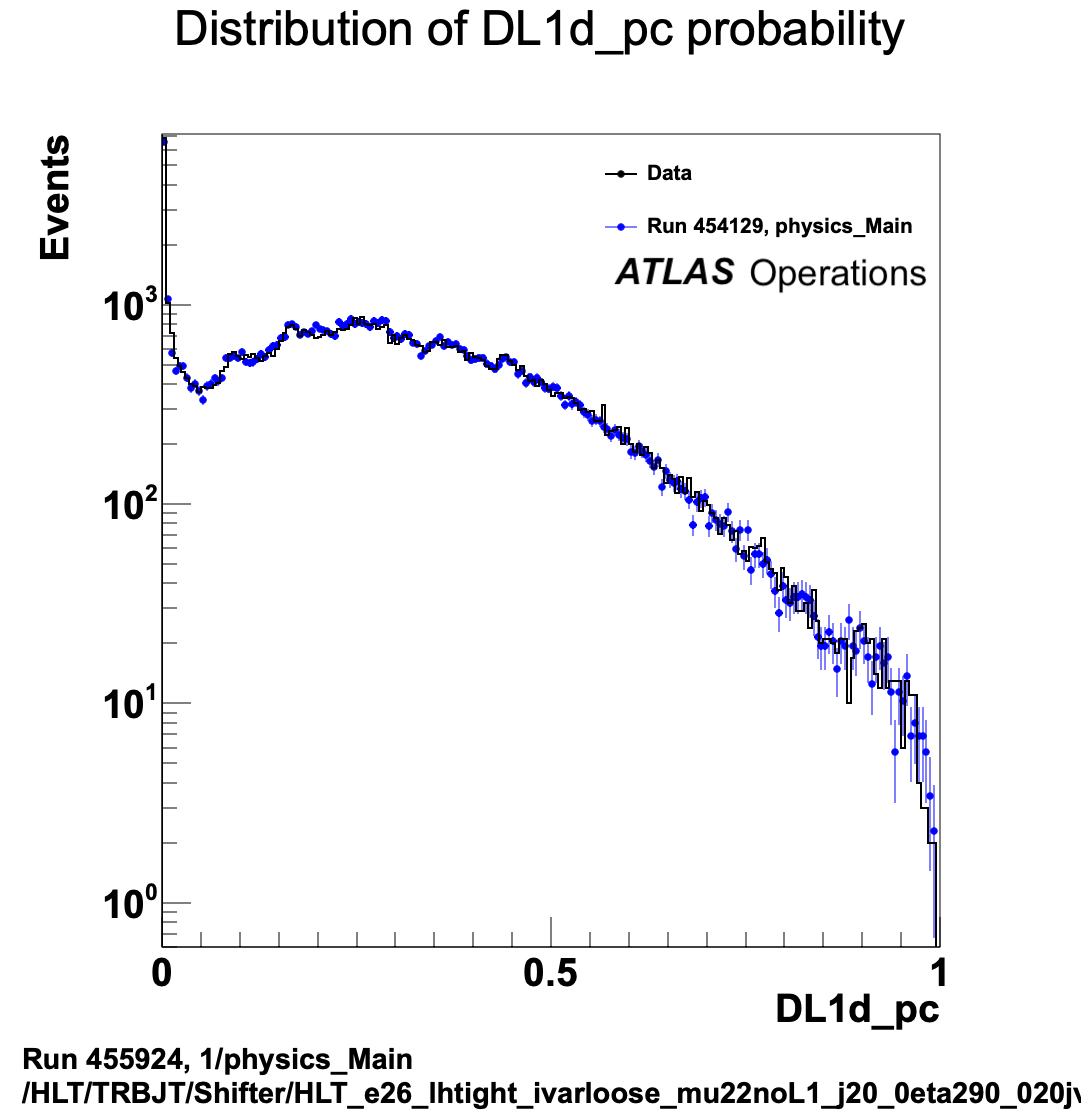}}}\hfill \\
\subfloat[]{\adjustbox{width=.44\columnwidth, cfbox=Gold 3pt 2pt}{\includegraphics{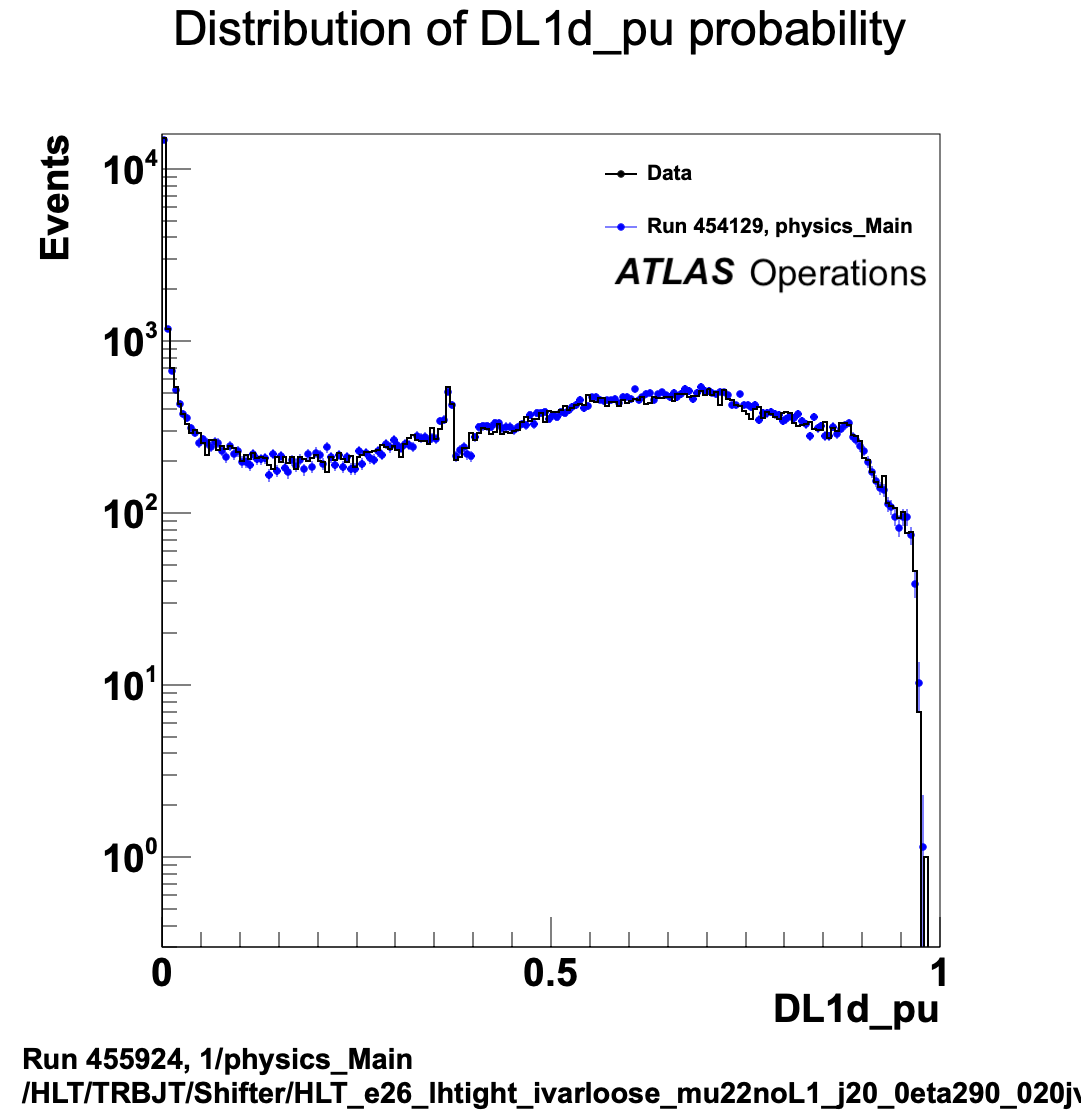}}}\hfill
\subfloat[]{\adjustbox{width=.44\columnwidth, cfbox=OliveDrab 3pt 2pt}{\includegraphics{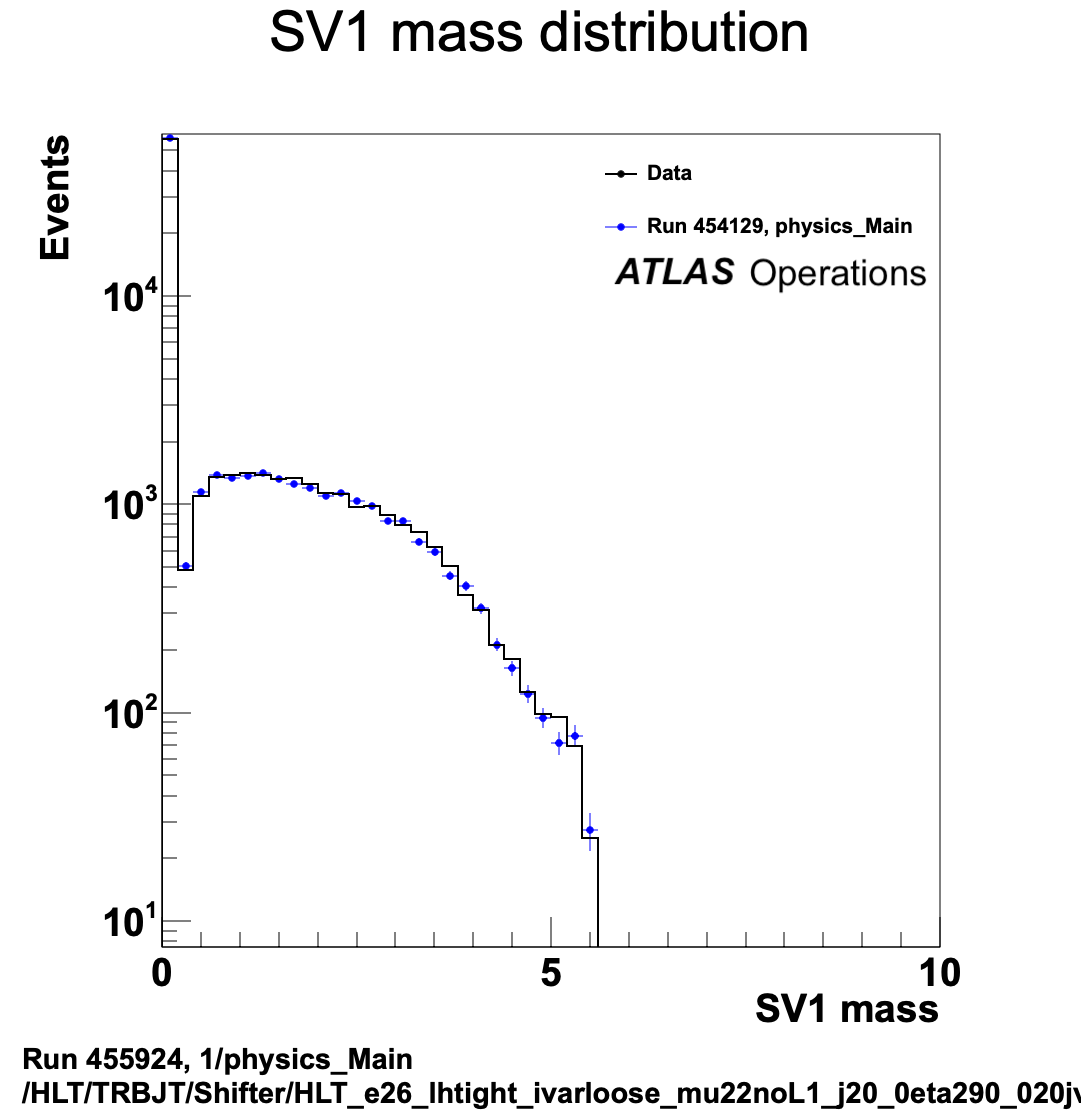}}} \\
\caption{Examples of the monitored quantities present in the web display for the offline data quality assessment. Shown here are various plots for $b$-jet candidates as reconstructed in the HLT.
The display
shows $b$-tagging quantities for dedicated dilepton plus jets chains, enhanced in $t\bar{t}$ events, the \dloned probability for the (a) $b$-, (b) $c$-, (c) light-jet hypothesis and (d) the mass of the reconstructed Secondary Vertex (in GeV).
The data for the relevant run (solid histogram) are compared with a reference (solid markers), and
according to a $\chi^2$ test, depending on the agreement, they are flagged as green ($\chi^2/\textrm{Ndof}<1$), yellow ($1<\chi^2/\textrm{Ndof}<1.5$), or red ($\chi^2/\textrm{Ndof}>1.5$). The reference run is updated periodically to reflect significant condition changes, such as updated trigger menus. This example shows run 455924, which was recorded in April 2023. Error bars include the statistical uncertainties only.}
\label{fig:mon}
\end{figure}

By using the DQMD in the ATLAS control room, and the web-based display, the tasks of both online and offline shifters are facilitated by  a series of automatic tests, such as width and mean of the distribution comparison, or the Kolmogorov-Smirnov statistical test, which compares several parameters of a distribution to a reference histogram to evaluate data quality.
Histograms then appear in different colours depending on to which degree the shown distributions satisfy these tests, for instance a $\chi^2$ test, depending on the agreement, flags as green if the ratio of $\chi^2$ over the number of degrees of freeedom ($\chi^2/\textrm{Ndof}$) is smaller than one, yellow if $1<\chi^2/\textrm{Ndof}<1.5$ and finally red if $\chi^2/\textrm{Ndof}>1.5$. This is shown in Figure~\ref{fig:mon} where several $b$-tagging related quantities for a dilepton(one electron and one muon)-plus jets chain, naturally enhanced in $t\bar{t}$ events, for the run 455924, recorded in April 2023, are compared with a reference.
Overall, mismatches can be due to actual problems such as differences in the run pile-up profile and instantaneous luminosity, slight differences in the beam position, LHC configuration changes, up to detector or reconstruction problems and are followed up and discussed in dedicated ATLAS forums.


%
%
\FloatBarrier


\section{Conclusion}

ATLAS restarted  its data taking in 2022 deploying new state-of-the-art $b$-tagging algorithms constructed from neural networks.
In the first year of the Run 3 data-taking the new \dloned algorithm led already to a sizeable increase in performance relative to Run~2.
For the 2023 data-taking campaign the baseline tagger changed to the \gnone algorithm which added up to a factor two improvement in light-flavour rejection.
These improvements are introduced together with a new two-step trigger strategy that mitigates the CPU demands by introducing a fast $b$-tagging preselection, running on calorimeter-based jets, which is scheduled before the HLT precision-tracking-based $b$-tagging. These ameliorations allow ATLAS to collect hadronic signatures such as \hhbbtt with an increase in signal efficiency of up to a factor 1.7 relative to Run 2. %
Finally, to assess the quality of the simulation, conditional efficiencies are measured in data and compared with simulation in $t\bar{t}$-enriched events, where an overall good agreement
of the trigger efficiencies are observed.


%
\section*{Acknowledgements}
%

%
%

%
%

We thank CERN for the very successful operation of the LHC and its injectors, as well as the support staff at
CERN and at our institutions worldwide without whom ATLAS could not be operated efficiently.

The crucial computing support from all WLCG partners is acknowledged gratefully, in particular from CERN, the ATLAS Tier-1 facilities at TRIUMF/SFU (Canada), NDGF (Denmark, Norway, Sweden), CC-IN2P3 (France), KIT/GridKA (Germany), INFN-CNAF (Italy), NL-T1 (Netherlands), PIC (Spain), RAL (UK) and BNL (USA), the Tier-2 facilities worldwide and large non-WLCG resource providers. Major contributors of computing resources are listed in Ref.~\cite{ATL-SOFT-PUB-2023-001}.

We gratefully acknowledge the support of ANPCyT, Argentina; YerPhI, Armenia; ARC, Australia; BMWFW and FWF, Austria; ANAS, Azerbaijan; CNPq and FAPESP, Brazil; NSERC, NRC and CFI, Canada; CERN; ANID, Chile; CAS, MOST and NSFC, China; Minciencias, Colombia; MEYS CR, Czech Republic; DNRF and DNSRC, Denmark; IN2P3-CNRS and CEA-DRF/IRFU, France; SRNSFG, Georgia; BMBF, HGF and MPG, Germany; GSRI, Greece; RGC and Hong Kong SAR, China; ICHEP and Academy of Sciences and Humanities, Israel; INFN, Italy; MEXT and JSPS, Japan; CNRST, Morocco; NWO, Netherlands; RCN, Norway; MNiSW, Poland; FCT, Portugal; MNE/IFA, Romania; MSTDI, Serbia; MSSR, Slovakia; ARIS and MVZI, Slovenia; DSI/NRF, South Africa; MICIU/AEI, Spain; SRC and Wallenberg Foundation, Sweden; SERI, SNSF and Cantons of Bern and Geneva, Switzerland; NSTC, Taipei; TENMAK, T\"urkiye; STFC/UKRI, United Kingdom; DOE and NSF, United States of America.

Individual groups and members have received support from BCKDF, CANARIE, CRC and DRAC, Canada; CERN-CZ, FORTE and PRIMUS, Czech Republic; COST, ERC, ERDF, Horizon 2020, ICSC-NextGenerationEU and Marie Sk{\l}odowska-Curie Actions, European Union; Investissements d'Avenir Labex, Investissements d'Avenir Idex and ANR, France; DFG and AvH Foundation, Germany; Herakleitos, Thales and Aristeia programmes co-financed by EU-ESF and the Greek NSRF, Greece; BSF-NSF and MINERVA, Israel; NCN and NAWA, Poland; La Caixa Banking Foundation, CERCA Programme Generalitat de Catalunya and PROMETEO and GenT Programmes Generalitat Valenciana, Spain; G\"{o}ran Gustafssons Stiftelse, Sweden; The Royal Society and Leverhulme Trust, United Kingdom.

In addition, individual members wish to acknowledge support from Armenia: Yerevan Physics Institute (FAPERJ); CERN: European Organization for Nuclear Research (CERN DOCT); Chile: Agencia Nacional de Investigaci\'on y Desarrollo (FONDECYT 1230812, FONDECYT 1230987, FONDECYT 1240864); China: Chinese Ministry of Science and Technology (MOST-2023YFA1605700, MOST-2023YFA1609300), National Natural Science Foundation of China (NSFC - 12175119, NSFC 12275265, NSFC-12075060); Czech Republic: Czech Science Foundation (GACR - 24-11373S), Ministry of Education Youth and Sports (FORTE CZ.02.01.01/00/22\_008/0004632), PRIMUS Research Programme (PRIMUS/21/SCI/017); EU: H2020 European Research Council (ERC - 101002463); European Union: European Research Council (ERC - 948254, ERC 101089007, ERC, BARD, 101116429), European Union, Future Artificial Intelligence Research (FAIR-NextGenerationEU PE00000013), Italian Center for High Performance Computing, Big Data and Quantum Computing (ICSC, NextGenerationEU); France: Agence Nationale de la Recherche (ANR-20-CE31-0013, ANR-21-CE31-0013, ANR-21-CE31-0022, ANR-22-EDIR-0002); Germany: Baden-Württemberg Stiftung (BW Stiftung-Postdoc Eliteprogramme), Deutsche Forschungsgemeinschaft (DFG - 469666862, DFG - CR 312/5-2); Italy: Istituto Nazionale di Fisica Nucleare (ICSC, NextGenerationEU), Ministero dell'Università e della Ricerca (PRIN - 20223N7F8K - PNRR M4.C2.1.1); Japan: Japan Society for the Promotion of Science (JSPS KAKENHI JP22H01227, JSPS KAKENHI JP22H04944, JSPS KAKENHI JP22KK0227, JSPS KAKENHI JP23KK0245); Netherlands: Netherlands Organisation for Scientific Research (NWO Veni 2020 - VI.Veni.202.179); Norway: Research Council of Norway (RCN-314472); Poland: Ministry of Science and Higher Education (IDUB AGH, POB8, D4 no 9722), Polish National Agency for Academic Exchange (PPN/PPO/2020/1/00002/U/00001), Polish National Science Centre (NCN 2021/42/E/ST2/00350, NCN OPUS 2023/51/B/ST2/02507, NCN OPUS nr 2022/47/B/ST2/03059, NCN UMO-2019/34/E/ST2/00393, NCN \& H2020 MSCA 945339, UMO-2020/37/B/ST2/01043, UMO-2021/40/C/ST2/00187, UMO-2022/47/O/ST2/00148, UMO-2023/49/B/ST2/04085, UMO-2023/51/B/ST2/00920); Slovenia: Slovenian Research Agency (ARIS grant J1-3010); Spain: Generalitat Valenciana (Artemisa, FEDER, IDIFEDER/2018/048), Ministry of Science and Innovation (MCIN \& NextGenEU PCI2022-135018-2, MICIN \& FEDER PID2021-125273NB, RYC2019-028510-I, RYC2020-030254-I, RYC2021-031273-I, RYC2022-038164-I); Sweden: Carl Trygger Foundation (Carl Trygger Foundation CTS 22:2312), Swedish Research Council (Swedish Research Council 2023-04654, VR 2018-00482, VR 2022-03845, VR 2022-04683, VR 2023-03403, VR grant 2021-03651), Knut and Alice Wallenberg Foundation (KAW 2018.0458, KAW 2019.0447, KAW 2022.0358); Switzerland: Swiss National Science Foundation (SNSF - PCEFP2\_194658); United Kingdom: Leverhulme Trust (Leverhulme Trust RPG-2020-004), Royal Society (NIF-R1-231091); United States of America: U.S. Department of Energy (ECA DE-AC02-76SF00515), Neubauer Family Foundation.

%
%


%
%
%
%
%
%

%
%
%
%
%
%

%
%

%
%
%
\printbibliography
\clearpage
 
\begin{flushleft}
\hypersetup{urlcolor=black}
{\Large The ATLAS Collaboration}

\bigskip

\AtlasOrcid[0000-0002-6665-4934]{G.~Aad}$^\textrm{\scriptsize 104}$,
\AtlasOrcid[0000-0001-7616-1554]{E.~Aakvaag}$^\textrm{\scriptsize 17}$,
\AtlasOrcid[0000-0002-5888-2734]{B.~Abbott}$^\textrm{\scriptsize 123}$,
\AtlasOrcid[0000-0002-0287-5869]{S.~Abdelhameed}$^\textrm{\scriptsize 119a}$,
\AtlasOrcid[0000-0002-1002-1652]{K.~Abeling}$^\textrm{\scriptsize 56}$,
\AtlasOrcid[0000-0001-5763-2760]{N.J.~Abicht}$^\textrm{\scriptsize 50}$,
\AtlasOrcid[0000-0002-8496-9294]{S.H.~Abidi}$^\textrm{\scriptsize 30}$,
\AtlasOrcid[0009-0003-6578-220X]{M.~Aboelela}$^\textrm{\scriptsize 46}$,
\AtlasOrcid[0000-0002-9987-2292]{A.~Aboulhorma}$^\textrm{\scriptsize 36e}$,
\AtlasOrcid[0000-0001-5329-6640]{H.~Abramowicz}$^\textrm{\scriptsize 155}$
% ...
\bigskip
\\

% ...
$^{17}$Department for Physics and Technology, University of Bergen, Bergen; Norway.\\
% ...
$^{30}$Physics Department, Brookhaven National Laboratory, Upton NY; United States of America.\\
% ...
$^{36}$$^{(a)}$Facult\'e des Sciences Ain Chock, Universit\'e Hassan II de Casablanca;$^{(b)}$Facult\'{e} des Sciences, Universit\'{e} Ibn-Tofail, K\'{e}nitra;$^{(c)}$Facult\'e des Sciences Semlalia, Universit\'e Cadi Ayyad, LPHEA-Marrakech;$^{(d)}$LPMR, Facult\'e des Sciences, Universit\'e Mohamed Premier, Oujda;$^{(e)}$Facult\'e des sciences, Universit\'e Mohammed V, Rabat;$^{(f)}$Institute of Applied Physics, Mohammed VI Polytechnic University, Ben Guerir; Morocco.\\
% ...
$^{46}$Physics Department, Southern Methodist University, Dallas TX; United States of America.\\
% ...
$^{50}$Fakult\"{a}t Physik , Technische Universit{\"a}t Dortmund, Dortmund; Germany.\\
% ...
$^{56}$II. Physikalisches Institut, Georg-August-Universit\"{a}t G\"ottingen, G\"ottingen; Germany.\\
% ...
$^{104}$CPPM, Aix-Marseille Universit\'e, CNRS/IN2P3, Marseille; France.\\
% ...
$^{119}$$^{(a)}$New York University Abu Dhabi, Abu Dhabi;$^{(b)}$United Arab Emirates University, Al Ain; United Arab Emirates.\\
% ...
$^{123}$Homer L. Dodge Department of Physics and Astronomy, University of Oklahoma, Norman OK; United States of America.\\
% ...
$^{155}$Raymond and Beverly Sackler School of Physics and Astronomy, Tel Aviv University, Tel Aviv; Israel.\\
% ...
$^{a}$ Also Affiliated with an institute covered by a cooperation agreement with CERN.\\
$^{b}$ Also at An-Najah National University, Nablus; Palestine.\\
$^{c}$ Also at Borough of Manhattan Community College, City University of New York, New York NY; United States of America.\\
$^{d}$ Also at Center for High Energy Physics, Peking University; China.\\
$^{e}$ Also at Center for Interdisciplinary Research and Innovation (CIRI-AUTH), Thessaloniki; Greece.\\
$^{f}$ Also at CERN, Geneva; Switzerland.\\
% ...

\end{flushleft}


%

%
%
%
%

%
%
%
%

\end{document}